\documentclass[]{emulateapj}
\usepackage{apjfonts}
\usepackage{ulem}
\usepackage{threeparttable}
\usepackage{subfigure}
\usepackage{graphicx}
\usepackage{amsmath}
\usepackage{color}
\usepackage{appendix}
\usepackage{etoolbox}
\usepackage{natbib}
\usepackage{lineno}
\usepackage[breaklinks,colorlinks,citecolor=blue,linkcolor=blue]{hyperref}
\makeatletter
 \patchcmd{\NAT@citex}
 {\@citea\NAT@hyper@{%
 \NAT@nmfmt{\NAT@nm}%
 \hyper@natlinkbreak{\NAT@aysep\NAT@spacechar}{\@citeb\@extra@b@citeb}%
 \NAT@date}}
 {\@citea\NAT@nmfmt{\NAT@nm}%
 \NAT@aysep\NAT@spacechar\NAT@hyper@{\NAT@date}}{}{}

 \patchcmd{\NAT@citex}
 {\@citea\NAT@hyper@{%
 \NAT@nmfmt{\NAT@nm}%
 \hyper@natlinkbreak{\NAT@spacechar\NAT@@open\if#1\else#1\NAT@spacechar\fi}%
 {\@citeb\@extra@b@citeb}%
 \NAT@date}}
 {\@citea\NAT@nmfmt{\NAT@nm}%
 \NAT@spacechar\NAT@@open\if#1\else#1\NAT@spacechar\fi\NAT@hyper@{\NAT@date}}
 {}{}
 \makeatother

\begin{document}
\newcommand{\AlIII}{Al\,{\footnotesize III}}
\newcommand{\AlII}{Al\,{\footnotesize II}}
\newcommand{\Lya}{Ly$\alpha$}
\newcommand{\Lyb}{Ly$\beta$}
\newcommand{\Lyg}{Ly$\gamma$}
\newcommand{\Lyd}{Ly$\delta$}
\newcommand{\Lye}{Ly$\epsilon$}
\newcommand{\NV}{\ion{N}{5}}
\newcommand{\SiIV}{\ion{Si}{4}}
\newcommand{\SiIII}{\ion{Si}{3}}
\newcommand{\SiII}{\ion{Si}{2}}
\newcommand{\CIV}{\ion{C}{4}}
\newcommand{\PV}{P\,{\footnotesize V}}
\newcommand{\SIV}{\ion{S}{4}}
\newcommand{\CIII}{C\,{\footnotesize III}]}
\newcommand{\CII}{C\,{\footnotesize II}}
\newcommand{\MgII}{Mg\,{\footnotesize II}}
\newcommand{\SII}{S\,{\footnotesize II}}
\newcommand{\SIII}{S\,{\footnotesize III}}
\newcommand{\HI}{\ion{H}{1}}
\newcommand{\Hr}{H$\gamma$}
\newcommand{\Hb}{H$\beta$}
\newcommand{\Ha}{H$\alpha$}
\newcommand{\pr}{Pa$\gamma$}
\newcommand{\OII}{[\ion{O}{2}]}
\newcommand{\OIII}{[\ion{O}{3}]}
\newcommand{\OVI}{\ion{O}{6}}
\newcommand{\FeII}{Fe\,{\footnotesize II}}
\newcommand{\FeIII}{Fe\,{\footnotesize III}}
\newcommand{\HeI}{He\,{\footnotesize I}}
\newcommand{\heii}{He\,{\footnotesize II}}
\newcommand{\NII}{N\,{\footnotesize II}}
\newcommand{\NIII}{\ion{N}{3}}
\newcommand{\sii}{[S\,{\footnotesize II}]}
\newcommand{\feii}{Fe\,{\footnotesize II}}
\newcommand{\NH}{$N_{\rm H}$}
\newcommand{\nH}{$n_{\rm H}$}
\newcommand{\NHU}{$N_{\rm H}-U$}
\newcommand{\kmps}{$\rm km~s^{-1}$}
\newcommand{\mbh}{$M_{\rm BH}$}
\newcommand{\Msun}{$M_{\odot}$}
\newcommand{\rielr}{$R_{\rm IELR}$}
\newcommand{\rsub}{$R_{\rm sub}$}
\newcommand{\red}{\bf\color{red}}
\newcommand{\blue}{\color{blue}}
\newcommand{\magenta}{\bf\color{magenta}}
\newcommand{\green}{\bf\color{green}}
\newcommand{\dd}{\red\sout}
\newcommand{\rev}{\color{magenta}}
\newcommand{\ddd}{\rev\sout}
\title{Metal-strong Inflows at the Outer-galactic-scale of a Quasar}
\author{Qiguo Tian\altaffilmark{1,2}, Lei Hao\altaffilmark{3}, Yipeng Zhou\altaffilmark{1,2,3,4,5}, Xiheng Shi\altaffilmark{1,2}, Tuo Ji\altaffilmark{1,2}, Peng Jiang\altaffilmark{1,2}, Lin Lin\altaffilmark{3}, Zhenya Zheng\altaffilmark{3}, Hongyan Zhou\altaffilmark{1,2,6}}
\altaffiltext{1}{Polar Research Institute of China, 451 Jinqiao Road, Pudong,
Shanghai 200136, People's Republic of China; {\blue tianqiguo@pric.org.cn, zhouhongyan@pric.org.cn }}
\altaffiltext{2}{MNR Key Laboratory for Polar Science, Polar Research Institute of China,
451 Jinqiao Road, Pudong, Shanghai 200136, People's Republic of China}
\altaffiltext{3}{Shanghai Astronomical Observatory, Chinese Academy of Sciences, 80 Nandan Road, Shanghai 200030, People's Republic of China}
\altaffiltext{4}{School of Astronomy and Space Sciences, University of Chinese Academy of Sciences, 19A Yuquan Road, Beijing 100049, People's Republic of China}
\altaffiltext{5}{School of Astronomy and Space Science, Nanjing University, Nanjing, Jiangsu 210093, People's Republic of China}
\altaffiltext{6}{Key laboratory for Research in Galaxies and Cosmology of
Chinese Academy of Science, Department of Astronomy, University of Science
and Technology of China, Hefei, Anhui 230026, People's Republic of China}

\begin{abstract}
We present an analysis of the absorption-line system in the 
Very Large Telescope/Ultraviolet and Visual Echelle Spectrograph spectrum
at a redshift of $z_{\rm a}={3.1448}$ associated with the quasar SDSS
J122040.23+092326.96, whose systematic redshift is $z_{\rm e}=3.1380\pm0.0007$,
measured from the ${\rm H}\beta$+{\OIII} emission lines in our newly acquired
NIR P200/TripleSpec data. This absorbing system, detected in numerous absorption lines
including the {\NV}, {\NIII}, {\CIV}, \ion{C}{3}, {\SiIV}, {\SiIII}, and {\HI} 
Lyman series, can be resolved into seven kinematic components with red-shifted 
velocities ranging from 200 to $900\,\rm km\,s^{-1}$. The high-ionization {\NV} 
doublet detected and the rather narrow Lyman series measured ($b\approx14\,\rm km\,s^{-1}$) 
suggest that the absorption gas is photo ionized, possibly by the quasar. A low 
density is inferred by the fact that {\NIII} $\lambda989.80$ is significantly 
detected while {\NIII*} $\lambda991.51$ (${\rm log}\,n_{\rm c}=3.3\,\rm cm^{-3}$) 
is undetectably weak. A firm lower limit of a solar value to the abundance of the
gas can be set based on the measurements of {\SiIV} and {\HI} column densities,
as first proposed by F. Hamann. Detailed photoionization simulations
indicate that $T1$, and possibly the absorber as a whole, has metallicities
of $Z\sim1.5-6.0\,Z_\sun$, and is located at $\sim15\,\rm kpc$ from the quasar
nucleus. The metal-strong absorption inflows at the outskirt of the quasar host
galaxy is most likely originated in situ and were driven by stellar processes, such
as stellar winds and/or supernova explosions. Such a relatively rare system may 
hold important clues to understanding the baryonic cycling of galaxies, and more 
cases could be picked out using relatively strong {\SiIV} and weak Lyman 
absorption lines.

\end{abstract}


\section{Introduction}
Chemical evolution, i.e., the enrichment, dilution, and cycling of elements
heavier than H and He, is one of the fundamental processes involved in
galaxy evolution \citep[e.g.,][]{Dwek1998,Elvis2006,Borguet2012a,Arcones2023}.
Galaxies accrete the metal-poor gas from the intergalactic medium (IGM),
which is then transformed into stars via star formation processes. The
heavy elements are synthesized in the cores of stars, and with the death
of these stars, these elements are further redistributed throughout the
galaxies along with the baryonic cycling of galaxies. Outflows from
galaxies can transport heavy metals over large distances, thereby enhancing
the abundance of the circumgalactic medium or the IGM. Part of
these gases can later be re-accreted onto the galaxy, providing raw
materials for the next generation of formation of stars and planets
\citep[e.g.,][]{Kobayashi2020}.  These processes constitute the framework
that regulates the metal content of the galaxy and help shape the
fundamental relationships involving the metallicities of galaxies.
Therefore, studies on the metallicity of galaxies encode information
about the star formation history of the galaxy and simultaneously
provide insights into the complex processes that regulate the metal
content, such as the large-scale gas inflows and outflows.

Quasars are extraordinarily luminous objects  powered by the accretion of gas 
onto supermassive black holes \citep[BHs; e.g.,][]{Netzer2015}. There is 
growing evidence that metal-rich outflows from quasars are important sources 
of metal enrichment to the interstellar medium (ISM)/IGM (e.g., F. Hamann et al. \citeyear{Hamann2002}; 
V. D'Odorico et al. \citeyear{DOdorico2004a}, \citeyear{DOdorico2004b}; 
S. Veilleux et al. \citeyear{Veilleux2005}; S. K. Ballero et al. \citeyear{Ballero2008}; 
B. C. J. Borguet et al. \citeyear{Borguet2012a}; S. Lai et al. \citeyear{Lai2022}). Indeed,
observations by Chandra \citep{Nicastro2005} have shown that the IGM has been enriched
with heavy elements, far from having a primordial composition \citep{Fang2002,Pettini2004}.

Quasars often exhibit abundant absorption and emission lines \citep{Weymann1991},
which are valuable tools for drawing information on chemical abundances
($Z$) \citep[e.g.,][]{Borguet2012a}. Compared with emission lines, absorption
lines have certain advantages in abundance determinations, for example, they provide
diagnostics that largely do not depend on the detailed understanding of the temperature 
and density \citep{Hamann1998}. If the associated absorbing gas happens to be located 
at large distance from the galaxy center (e.g., $>10\,\rm kpc$), the metallicity inferred 
can be a unique probe to determine the properties of the host galaxy and its 
environment \citep{Hamann2004,Veilleux2005,Tumlinson2017,Chen2019}.

According to their velocity ranges, the absorption-line systems in quasars can 
be empirically divided into three categories: broad absorption lines (BALs) with 
a typical velocity range exceeding $2000\,\rm km\,s^{-1}$; narrow absorption
lines (NALs) with a velocity range less than $500\,\rm km\,s^{-1}$; and mini-BALs with
intermediate velocity range in between the above two types \citep{Hamann2004}.
It is normally believed that intrinsic NALs, mini-BALs, and BALs arise from
different components of absorbing gas, or different regions of the same
component \citep{Elvis2000,Misawa2007,Lu2008}. Based on the type of observed
absorption features, BALs, for example, are classified into high-ionization
BALs (HiBALs) and low-ionization BALs \citep[LoBALs; e.g.,][]{Weymann1991, Hall2002}.
Generally, HiBAL quasar spectra exhibit absorption lines of
{\OVI} $\lambda\lambda1031.91$, 1037.61, {\CIV} $\lambda\lambda1548.19$,
1550.77, {\NV} $\lambda\lambda1238.82$, 1242.80,  {\SiIV} $\lambda\lambda1393.76$,
1402.77 doublets, and {\HI} Lyman series \citep[e.g.,][]{Borguet2012a, Jiang2023}. 
Meanwhile, LoBAL spectra show not only HiBALs, but also
the absorption lines of {\MgII} $\lambda\lambda2796.35$, 2803.53,
{\AlIII} $\lambda\lambda1854.72$, 1862.79, {\CII}, {\SiII} and {\NIII}, etc.

Exploring metal abundance using absorption lines is achieved by comparing
observed and photoionization-simulated ion column densities of various
elements in a 4D-parameter space involving the hydrogen column density
({\NH}), hydrogen number density ({\nH}), ionization parameter ($U$),
and $Z$. For convenience, $Z$ is defined to be the scaled value of
the solar abundance. However, finding the best model in a 4D-parameter
space ({\NH}, {\nH}, $U$, and $Z$) that can simultaneously reproduce the
observed column densities of several ions is challenging.

F. Hamann (\citeyear{Hamann1997}) suggested that lower limits on metallicities can
be estimated using the combined information of {\HI} Lyman series and certain
high-ionization absorption lines. The values of the four parameters, including the
$Z$ are often determined iteratively by comparing the repeatedly refined model
results with the observed line strengths \citep[e.g.,][]{Arav2013, Borguet2013,Liu2016,Tian2019,Xu2019,Tian2021}.
As shown in Appendix \ref{appendix_A}, assuming that the spectral resolution and
signal-to-noise ratio (S/N) are sufficiently high, the column density of {\HI}
can be measured over a wide range, from $\sim10^{12}-10^{13}\,\rm cm^{-2}$ to
more than $10^{20}\,\rm cm^{-2}$, using Lyman absorption lines, Lyman-limit
system, and damped {\Lya} wing.

Among the high-ionization absorption lines, the {\OVI} $\lambda\lambda1031.91,\,1037.61$
and {\CIV} $\lambda\lambda1548.19,\,1550.77$ absorption lines, normally seen
in quasar spectra, are usually saturated, as oxygen and carbon are the two most
abundant metallic elements in solar abundance \citep[e.g.,][]{Allende2002}. In
addition, {\OVI} absorption doublet lines could be contaminated by {\Lya} forest, and
{\CIV} absorption doublet lines are often mixed together due to the velocity separation
being too small ($\sim500\,\rm km^{-1}$). In contrast, {\SiIV} absorption doublet
has some special advantages for measuring its column density: (1) {\SiIV} does
not suffer from the self-blending problem for NALs, and to some extent for mini-BALs,
since the velocity separation for {\SiIV} $\lambda\lambda1393.76,\,1402.77$ is
$\sim1933\rm\,km^{-1}$, which is much larger than that of {\CIV} doublet, (2)
the abundance of Si is lower than that of O and C by about 1 order
of magnitude \citep[e.g.,][]{Grevesse1998,Holweger2001,Allende2002}, thus, the
saturation problem for the {\SiIV} absorption lines is much weaker, (3)
absorption lines associated with other ions of Si element,
such as {\SiIII} {$\lambda1206.50$}, {\SiII} $\lambda\lambda1260.42,\,1526.71,\,1808.01$
and {\SiII*} $\lambda\lambda 1264.76,\,1533.43,\,1816.93$, if they could be detected,
can also be used jointly with {\SiIV} for diagnostics, as they a cover wide range
in both the wavelength and oscillator strengths.

In the present work, we report the detection of the metal-strong inflows located at
$\sim 15\,\rm kpc$ from the center BH of the quasar SDSS 122040.23+092326.96
(hereafter SDSS J1220+0923), using absorption lines, such as the {\HI} Lyman series,
{\SiIV}, {\NV}, etc., obtained from the Ultraviolet and Visual Echelle
Spectrograph (UVES) at the ESO's $8\rm\,m$ Very Large Telescope (VLT). An empirical
formula for estimating the lower limit to metallicity is also given.
This paper is organized as follows. The data we used will be described in
Section \ref{secobs}. We analyze the absorption lines and calculate the
corresponding ionic column densities in Section \ref{secspec}.
We compare the observational results with photoionization calculations in Section
\ref{secphys} and discuss the results in Section \ref{secdiss}. Our main findings
are summarized in Section \ref{secsumm}, along with the implications and future
prospects.

\section{Observations}\label{secobs}
\begin{figure*}[htb]
\center{\includegraphics[width=17cm]  {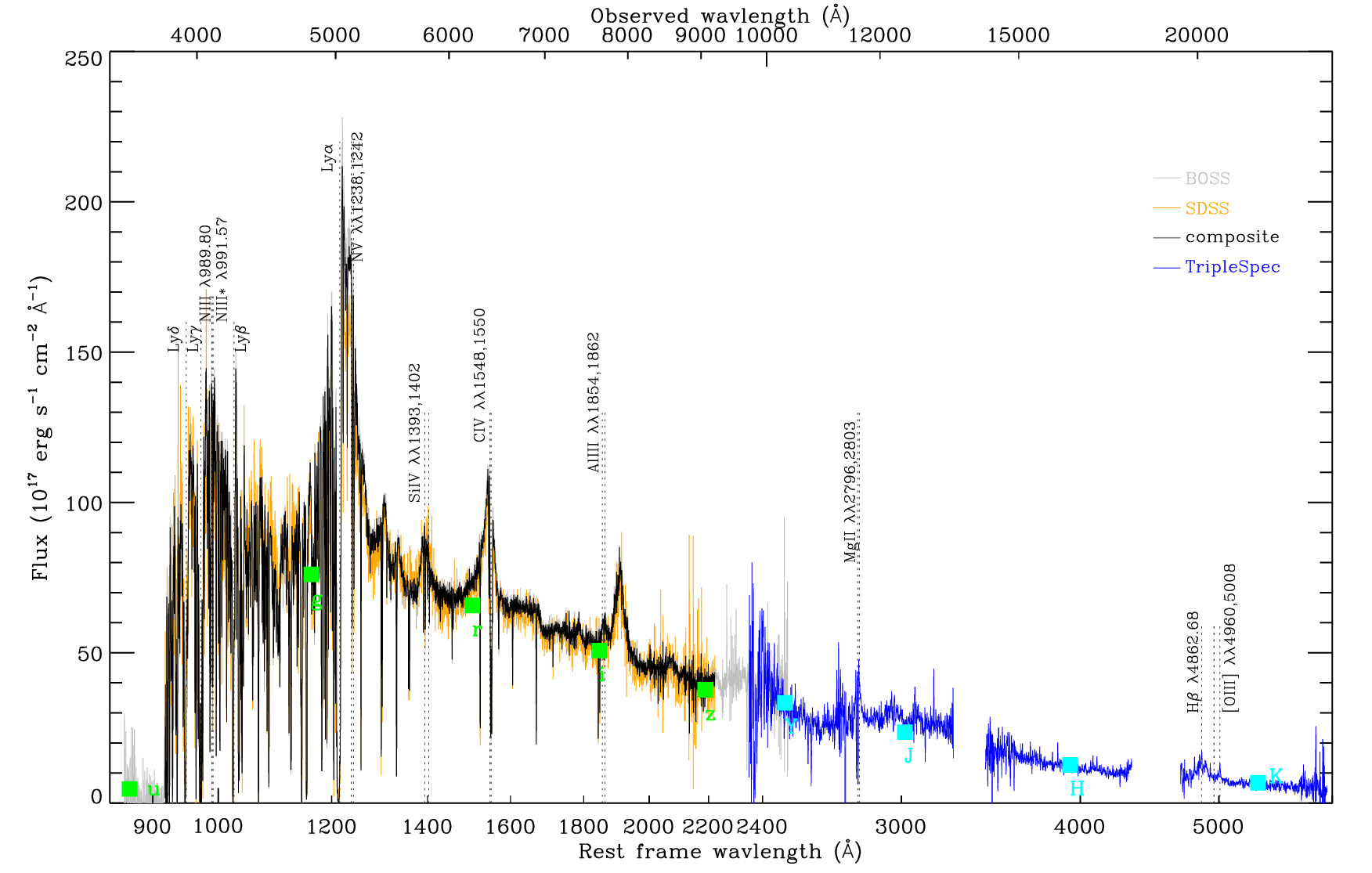}}
\caption{Rest-frame spectra of SDSS J1220+0923 from SDSS, BOSS, and TripleSpec.
The green- and cyan-filled squares are photometric data from the SDSS and UKIDSS surveys.
The TripleSpec spectrum (blue) presented has been smoothed for illustration
purposes.
}
\label{figspec}
\end{figure*}
SDSS J1220+0923 is a bright quasar
discovered in Sloan Digital Sky Survey \citep[SDSS;][]{Prochaska2008}.
The first spectrum was obtained with the SDSS spectrograph on 2003 February 2
\citep{SDSS2009}. The spectrum covers a wavelength range from 3800 to 9200 {\AA}
in the observed frame. The second spectrum was taken on 2012 January 21, using
the Baryon Oscillation Spectroscopic Survey (BOSS) spectrograph \citep{BOSS2013},
providing an extended wavelength coverage of $\lambda\sim3570-10350$ {\AA} in the 
observed frame. These two spectra share a similar spectral resolution of $R\sim2000$. 
The SDSS broadband photometry was carried out on 2003 April 27 \citep{York2000}. 
The magnitudes in the $g$, $r$, and $i$ bands agree well with those of the synthetic
photometry data obtained from the SDSS spectrum. The photometric monitoring
of this object by the Catalina Survey\footnote{The Catalina surveys website site is http://nesssi.cacr.caltech.edu/DataRelease/}
(for 8 yr since 2005 April 6) obtained 419 observations, showing weak long-term
variability of no more than $0.3\rm\,mag$ in the $V$ band. We thus use
the SDSS spectrum to recalibrate the BOSS spectrum, which is reported
to suffer from a serious flux-calibration issue \citep{BOSS2013}. Assuming
there are no prominent time variances of the quasar, we constructed a composite
spectrum by combining the SDSS and BOSS spectra weighted according to their spectral
S/Ns, which will be used for our further analysis.

\begin{figure}[htb]
\center{\includegraphics[width=8.6cm]  {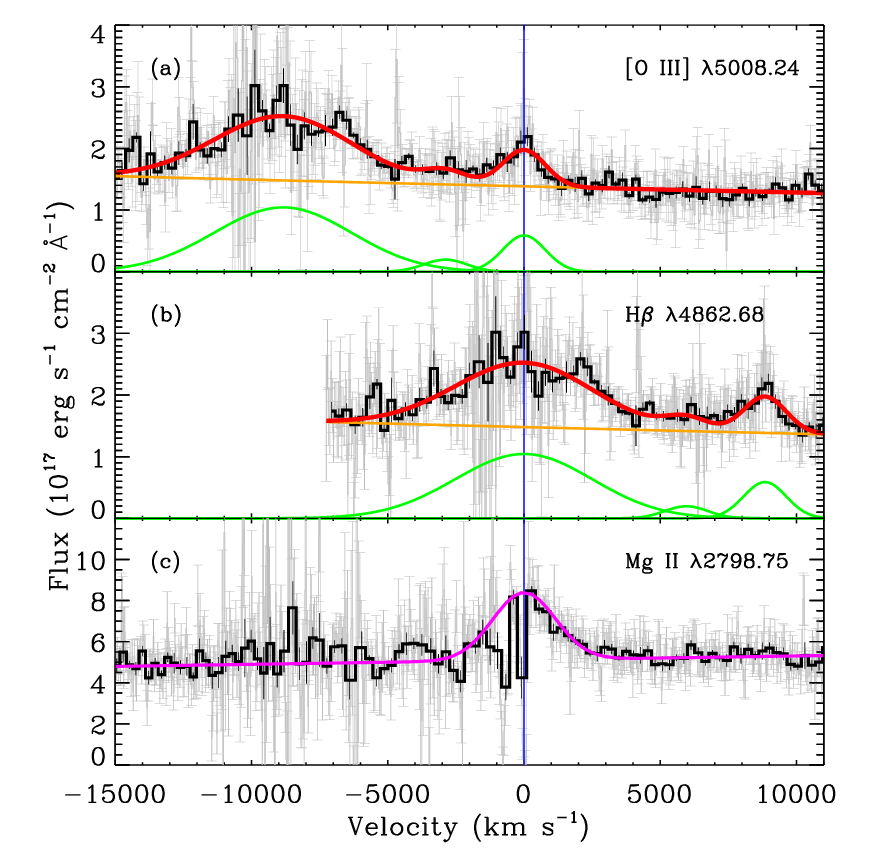}}
\caption{The TripleSpec spectrum of SDSS J1220+0923 showing (a) [\ion{O}{3}]$\,\lambda5008.24$,
(b) $\rm H\beta\,\lambda4862.68$, and (c) \ion{Mg}{2}$\,\lambda2798.75$. The gray line is the
observed spectrum, and the black line is the result rebinned using a weighted average of 
six data points for illustration purposes. The orange line in panels (a) and (b) represents 
the fitted power-law continuum in $\rm H\beta$ and [\ion{O}{3}] region, the three green lines are the fitted
Gaussian profiles for $\rm H\beta$ and [\ion{O}{3}] emissions, and the red line is the total 
fitting result. The common velocity space is based on the redshift measured from {\Hb} and [\ion{O}{3}].
The magenta line in panel (c) shows a Gaussian profile centered at $0\,\rm km\,s^{-1}$, with
the red wing matching the observed \ion{Mg}{2} spectrum (the blue wing is contaminated), indicating that \ion{Mg}{2} is consistent with
the measured redshift.
}
\label{figredshift}
\end{figure}

A follow-up near-infrared (NIR) spectroscopic observation was carried out on
2023  April 9 at the Hale 200 inch telescope, using the TripleSpec spectrograph
with a spectral resolution from 2500 to 2700. The NIR spectrum covers the rest-frame 
optical band (including the {\MgII} and {\Hb} regions). Four
exposures were taken, each for $300\,\rm s$. The seeing was about $1^{\prime\prime}$ during
the observations. The data was reduced with SpexTool \citep{Cushing2004}. The
flux calibration and telluric correction were performed with an IDL program
using the methods described in W. D. Vacca et al. (\citeyear{Vacca2003}). The extracted
flux agrees well with the United Kingdom Infrared Telescope (UKIRT) Infrared Deep
Sky Survey (UKIDSS) $YJHK$ photometric data obtained in 2007 February 7.\footnote{The
UKIDSS project is defined in A. Lawrence et al. (\citeyear{Lawrence2007}). UKIDSS uses
the UKIRT Wide Field Camera (WFCAM; \citealp{Casali2007}). The photometric system
is described in P. C. Hewett et al. (\citeyear{Hewett2006}), and the calibration is
described in S. T. Hodgkin et al. (\citeyear{Hodgkin2009}). The pipeline processing
and science archive are described by N. C. Hambly et al. (\citeyear{Hambly2008}).} The
NIR spectrum was used to calculate the systemic redshift, as it contains the 
{\OIII}$\,\lambda\lambda4960.30,\,5008.24$, and {\Hb}$\,\lambda4862.68$ emission
lines, with significances for {\Hb} and {\OIII} of about $13.3\sigma$ and $8.8\sigma$,
and a median spectral S/N in the {\Hb} and {\OIII} region of about 5.5. It 
has been suggested that emission lines such as the {\OII} $\lambda3728.48$,
{\OIII}$\,\lambda\lambda4960.30,\,5008.24$ and {\Hb}$\,\lambda4862.68$ can 
give good measurements of the systemic redshift \citep{Hewett2010,Shen2016}.
We jointly fitted the NIR spectrum covering the {\Hb} and {\OIII} region, 
using the \texttt{mpfitexpr} procedure in IDL, to measure
the systemic redshift. Each of the three lines was fitted by one Gaussian
profile, and the flux ratio of {\OIII}$\,\lambda5008.24/\lambda4960.30$ was
fixed to the theoretical value of 2.98 \citep[e.g.,][]{Storey2000,Dimitrijevi2007}.
The fitting result suggests a systemic redshift of $3.1380\pm0.0007$, which is close to
that observed by SDSS spectrum ($\sim3.1400$) \citep{Paris2018}, and it
will be used for further analysis in this work. After correcting for the
Galactic reddening with $E(B-V)=0.025\,\rm mag$ \citep{Schlegel1998,Fitzpatrick2007},
we transformed the photometric data, SDSS, BOSS, and composite spectra into the
rest frame with the newly derived systemic redshift, as shown in Figure \ref{figspec}.
Figure \ref{figredshift} shows the TripleSpec spectrum of {\Hb}, {\OIII}, and {\MgII} emission lines 
along with the fitted results. The average rest-wavelength of {\MgII} emission doublet 
(2798.75 {\AA}) is taken from D. E. Vanden Berk et al. (\citeyear{Vanden2001}), which is also
applied by Y. Shen et al. (\citeyear{Shen2016}) to measure the quasar's redshifts.
The {\MgII} emission line, as shown in Figure \ref{figredshift}(c), is consistent 
with the redshift determined from {\Hb} and {\OIII}, further validating our redshift measurement.

In addition, UVES Spectral Quasar Absorption Database (SQUAD)
released the high-resolution spectroscopic observations for SDSS
J1220+0923 \citep{Murphy2019} carried out using
UVES \citep{Dekker2000, Bagnulo2003}
at European Southern Observatory's (ESO's) $8\rm\,m$ VLT
on 2018 May 2.\footnote{SQUAD is a database of 467 high-resolution
quasar spectra from the data archive of VLT/UVES. The UVES spectroscopic
data can be found at https://data-portal.hpc.swin.edu.au/dataset/uves-squad-dr1.}
The exposure time for the observation was $12000\,\rm s$ and the spectroscopic resolution
was $R\sim40,000$. The released high-resolution normalized spectrum, covering
a wavelength range of $\lambda\sim3770-9467$ {\AA} in the observed frame,  will
be the backbone data, allowing us to probe the properties of the absorbing gas in SDSS J1220+0923.

\section{Spectral Analysis} \label{secspec}
\subsection{Detected Absorption Lines} \label{secdetect}
From the normalized high-spectral-resolution UVES spectrum, as shown
in Figure {\ref{figSiIVNV}}, we identified the high-ionization
absorption-line troughs of {\SiIV} $\lambda\lambda1393.76,1402.77$,
{\NV} $\lambda\lambda1238.82,1242.80$, and {\CIV} $\lambda1548.19, 1550.77$;
the low-ionization absorption troughs of {\NIII} $\lambda989.80$, and 
\ion{C}{3} $\lambda977.02$; and the absorption troughs of {\HI} Lyman series.
The absorption system can be roughly divided into seven components
($T1$ to $T7$) within which the absorption is continuous. Each component
contains one or several sub-kinematic components. The component $T1$,
which has the smallest velocity, is the most clearly detached trough.
The observed redshifts of the detected seven absorption components
are from 3.1411 to 3.1497, as shown in Table \ref{tabNcol}, which are
larger than the systemic redshift ($\sim3.1380$, see Section \ref{secobs}),
indicating that the absorption systems are redshifted. In this section, 
we will analyze the absorption lines and derive the column densities of 
the observed species. In the next section, these column densities will 
be used to derive the physical conditions of the inflowing gas.

\subsection{Column Density Measurements}

\subsubsection{{\NV} and {\SiIV}}\label{secNcolSiIV}
\begin{figure*}[htb]
\center{\includegraphics[width=18cm]  {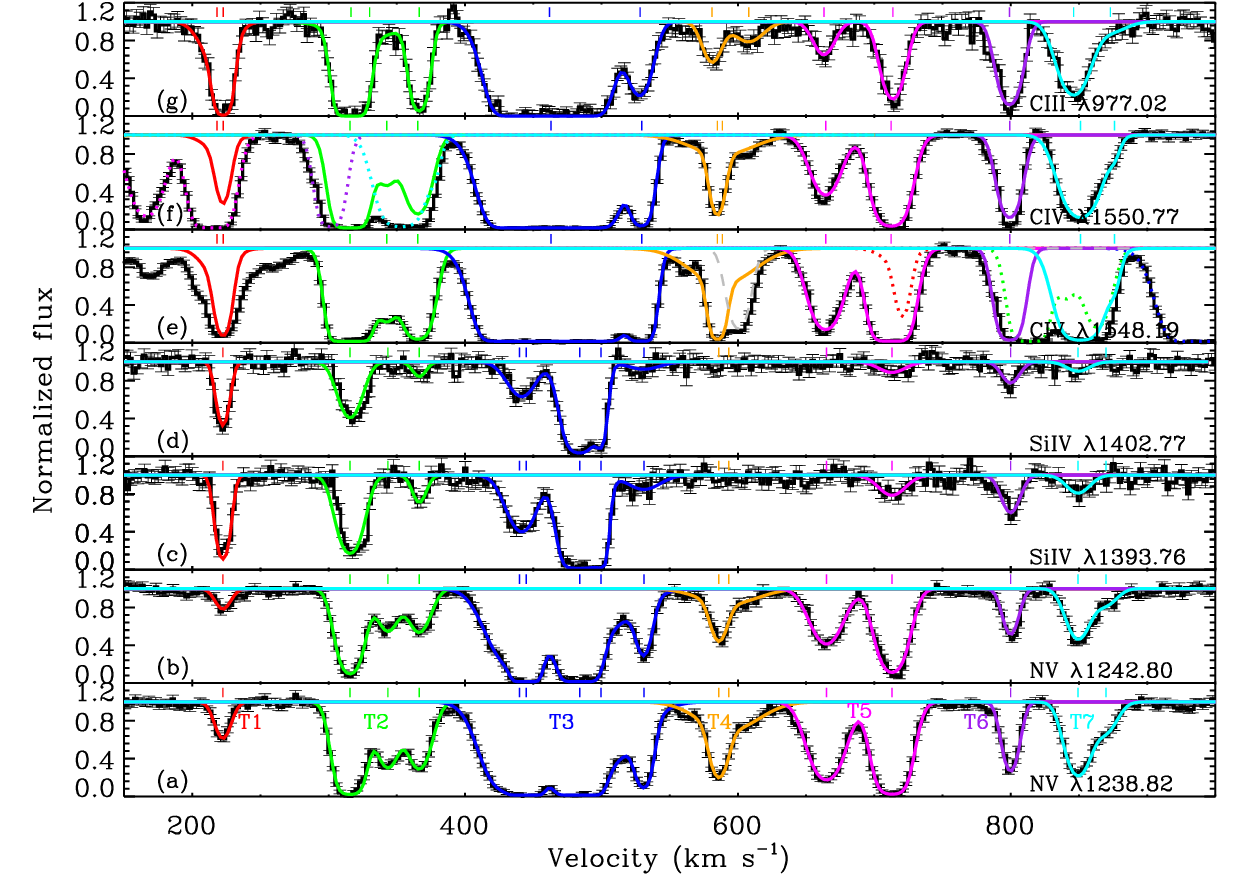}}
\caption{The absorption troughs of (a) {\NV} $\lambda1238.82$, (b) {\NV} $\lambda1242.80$,
(c) {\SiIV} $\lambda1393.76$, (d) {\SiIV} $\lambda1402.77$, (e) {\CIV} $\lambda1548.19$, 
(f) {\CIV} $\lambda1550.77$, and (g) \ion{C}{3} $\lambda977.02$. The seven components, 
fitted using Voigt functions, are presented in different colors with solid lines.
The blended features for {\CIV} $\lambda1548.19, 1550.77$ are presented with dotted
lines in panels (e) and (f); see the text for the details. The dashed gray line in panel (e) 
indicates the contamination in the component $T4$ for {\CIV} $\lambda1548.19$. The 
solid red lines in panels (e) and (f) represent the modeled upper-limit absorptions 
of {\CIV} $\lambda1548.19, 1550.77$ for component $T1$, based on the corresponding 
\ion{C}{3} $\lambda977.02$ profile for this component; see the text for the details. 
The small vertical lines in each panel indicate the centers of the fitted 
components.
}
\label{figSiIVNV}
\end{figure*}

Figure {\ref{figSiIVNV}}(a)--(d) shows the {\NV} $\lambda\lambda1238.82,\,1242.80$
and {\SiIV} $\lambda\lambda1393.76,\,1402.77$ absorption lines. We label
the seven sets of absorption trough components
with different colors in Figure \ref{figSiIVNV}(a). To obtain the column
densities of {\SiIV} and {\NV} (denoted as $N_{\rm Si\,IV}$ and $N_{\rm N\,V}$),
the rest-frame normalized spectrum covering the
absorption regions of {\SiIV} and {\NV} was jointly fitted  using Voigt
functions, with the \texttt{mpfitexpr} and \texttt{walvogit} procedures in IDL. 
For each of the seven components, a different number of Voigt
functions were used to characterize the optical depth $(\tau(v))$ for
the troughs of the {\NV} and {\SiIV} doublets. The components $T1$ and $T6$
use one Voigt function, $T4$, $T5$, and $T7$ use two, while $T2$ and $T3$
use three and five Voigt functions, respectively. The redshift of each
Voigt function associated with different transitions shares the same value.
Meanwhile, the widths of the {\SiIV} and {\NV} absorption lines are
considered as free parameters and are not required to be the same.
Then, the constructed optical depths for {\SiIV} and {\NV} were transformed
into a modeled normalized spectrum by the following equation:
\begin{equation}\label{eqtau}
I_r(v)=(1-C_f(v))+C_f(v)e^{-\tau(v)},
\end{equation}
where $C_f$ is the covering factor, which will be determined in the fitting.
We fit the model to the normalized spectrum, and the results are shown in
Figure \ref{figSiIVNV}, where the red ($T1$), green ($T2$), blue ($T3$),
orange ($T4$), magenta ($T5$), purple ($T6$), and cyan ($T7$) lines indicate
the seven absorption trough components. The $C_f$ was fitted to be $0.986\pm0.002$,
indicating that the {\SiIV} and {\NV} absorbing gas fully obscure the central
light source.

The ionic column densities can be obtained from the Voigt profile fitting, and
the relationship between ionic column densities and optical-depth profiles is
shown in Equation (\ref{eqNcol}) \citep[e.g.,][]{Arav2001},
\begin{equation}\label{eqNcol}
N_{\rm ion}=\frac{3.7679\times10^{14}}{f\lambda_0}\int\tau(v)\,dv\,(\rm cm^{-2}).
\end{equation}

The measured $N_{\rm Si\,IV}$ and $N_{\rm N\,V}$ are listed in Table \ref{tabNcol}.
Meanwhile, the Doppler $b$ parameter of the {\SiIV} absorption line for component
$T1$ is fitted to be $b_{\rm Si\,IV}=5.96\pm0.29\,{\rm km\,s^{-1}}$, which will
be used to estimate the gas temperature in Section \ref{secT} together with
$b_{\rm H\,I}$ estimated in Section \ref{secNcolHI}.

\subsubsection{ \ion{C}{3} and {\CIV}}
The \ion{C}{3} has an ionization potential of $47.89\,\rm eV$, which is
similar to that of {\SiIV} at $45.14\,\rm eV$. At the corresponding velocities,
\ion{C}{3} $\lambda977.02$ exhibits absorption components similar to those
observed in {\SiIV} and {\NV}, as shown in Figure \ref{figSiIVNV}(g).
Additionally, the ionization potential of {\CIV} is $64.49\,\rm eV$, which
lies between those of {\SiIV} and {\NV} ($97.89\,\rm eV$). Corresponding
absorption components have also been observed at the respective velocities
for {\CIV} $\lambda1548.19$ and {\CIV} $\lambda1550.77$, as shown in Figures
\ref{figSiIVNV}(e) and (f).

We apply the same number of Voigt functions to fit the \ion{C}{3} $\lambda977.02$
absorption features as we did for {\SiIV} and {\NV}, except for components $T3$
and $T1$. Since $T3$ is heavily saturated, we employ only two Voigt functions
(compared to five for {\SiIV} and {\NV}) to model its absorption. Meanwhile,
$T1$ has a stronger absorption in \ion{C}{3} compared to those of {\SiIV}
and {\NV} doublets (each fitted with one Voigt function); thus, we use two Voigt
functions to characterize its absorption feature. The fitted results for \ion{C}{3}
absorption lines are shown in Figure \ref{figSiIVNV}(g).

The velocity separation of the {\CIV} absorption doublet is $\sim500\rm\,km\,s^{-1}$,
which is smaller compared to those of {\NV} ($\sim960\rm\,km\,s^{-1}$) and {\SiIV}
($\sim1930\,\rm km\,s^{-1}$). Consequently, the 
blending of the {\CIV} absorptions, observed between transitions of different
systems, is due to the transitions of the doublet from the same system having such a
small velocity separation.
We use the same number of Voigt functions as for \ion{C}{3} $\lambda997.02$ to do
the fitting for the {\CIV} absorptions. It can be seen that the component $T1$ of
{\CIV} $\lambda1548.19$ (Figure \ref{figSiIVNV}(e)) has a very broad absorption
trough, indicating that it is likely contaminated. Therefore, we use the fitted
profile of \ion{C}{3} $\lambda977.02$ of $T1$ as a reference to give an upper
limit of $N_{\rm C\,IV}$ for this component. The fitted models are shown in
Figures \ref{figSiIVNV}(e) and (f) for {\CIV} $\lambda1548.19$
and {\CIV} $\lambda1550.77$.

The {\CIV} $\lambda1548.19$ absorption trough of $T5$ is blended with the
{\CIV} $\lambda1550.77$ absorption from $T1$, illustrated by the red dotted
line in Figure \ref{figSiIVNV}(e) for the contribution from {\CIV} $\lambda1550.77$
absorption. Similarly, the {\CIV} $\lambda1548.19$ absorption troughs of $T6$
and $T7$ are blended with the {\CIV} $\lambda1550.77$ absorption from $T2$,
as shown by the green dotted line in Figure \ref{figSiIVNV}(e), marking the
contribution from {\CIV} $\lambda1550.77$ absorption. In Figure \ref{figSiIVNV}(f),
we show the contribution of {\CIV} $\lambda1548.19$ absorption from $T5$ to
the {\CIV} $\lambda1550.77$ absorption trough of $T1$ with a magenta dotted line,
and show the contribution of {\CIV} $\lambda1548.19$ absorptions from $T6$
and $T7$ in the {\CIV} $\lambda1550.77$ absorption trough of $T2$ with purple
and cyan dotted lines. In addition, the component $T4$ for
{\CIV} $\lambda1548.19$ is contaminated, as indicated by the dashed gray
line in Figure \ref{figSiIVNV}(e). The measured column densities of
\ion{C}{3} and {\CIV} are listed in Table \ref{tabNcol}.

\subsubsection{Neutral Hydrogen}\label{secNcolHI}
\begin{figure*}[htb]
\center{\includegraphics[width=19cm]  {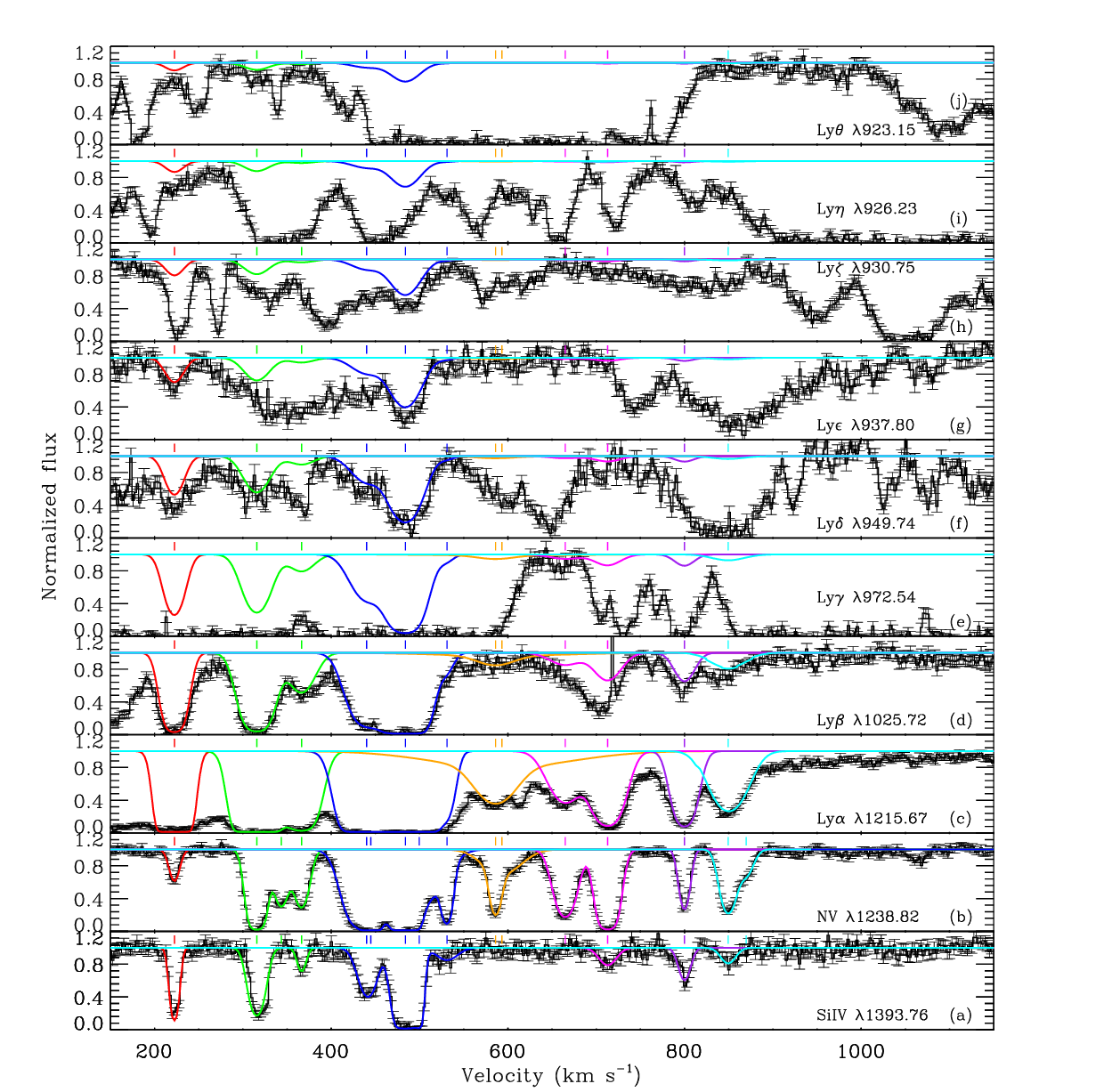}}
\caption{The normalized spectra in the \ion{Si}{4}, \ion{N}{5}, {\Lya}, {\Lyb},
Ly$\gamma$, Ly$\delta$, Ly$\epsilon$, Ly$\zeta$, Ly$\eta$, and Ly$\theta$ regions
with respect to \ion{Si}{4} $\lambda1393.76$, \ion{N}{5} $\lambda1238.82$,
{\Lya} $\lambda1215.67$, {\Lyb} $\lambda1025.73$, Ly$\gamma$ $\lambda972.54$,
Ly$\delta$ $\lambda949.74$, Ly$\epsilon$ $\lambda937.80$, Ly$\zeta$
$\lambda930.75$, Ly$\eta$ $\lambda926.23$, and Ly$\theta$ $\lambda923.15$.
The red, green, blue, orange, magenta, purple, and cyan lines represent the seven
absorption components. The small vertical lines in each panel
indicate the centers of the fitted components.
}
\label{fig:HI}
\end{figure*}
The normalized spectra of the {\HI} Lyman series from {\Lya} up to Ly$\theta$ are
presented in Figures \ref{fig:HI}(c)--(j). The velocity structures of
{\SiIV} $\lambda1393.76$ and {\NV} $\lambda1238.82$, from Figure \ref{figSiIVNV},
are also shown in Figures \ref{fig:HI}(a) and (b), to compare with those of the {\HI}
Lyman series.

We estimated the $N_{\rm H\,I}$ for each absorption component by shifting the
corresponding velocity centers of the modeled {\SiIV} $\lambda1393.76$
(and also {\NV} $\lambda1238.82$) profiles to those of the Lyman series while
adjusting the width and strength to match these absorption troughs. The
relative optical-depth values $\tau_i$ of {\HI} lines are set by the known
oscillator strengths.\footnote{$\tau_i=R\tau_{\rm Ly\alpha}$, where
$R=f_i\lambda_i/f_{\rm Ly\alpha}/\lambda_{\rm Ly\alpha}$ is the ratio of
the optical depth of the Lyman series to that of Ly$\alpha$, the $f_i$ are
the oscillator strengths, and the $\lambda_i$ are the wavelengths of the lines.}
The seven fitted absorption components are shown in red, green, blue, orange,
magenta, purple, and cyan, respectively, in Figure \ref{fig:HI}. Although there
are contaminations in the wavelength region of {\HI} Lyman series, possibly from
intervening {\HI} absorption lines \citep{Lynds1967,Weymann1991}, it is found
that, for all seven components, there are at least two modeled Lyman absorption
lines agree with the observations.  Taking component $T1$ as an example,
the modeled profiles can reproduce the troughs of {\Lyb} and {\Lye}
simultaneously, and those of Ly$\delta$, Ly$\eta$, and Ly$\theta$ are only
slightly shallower than the observations. Since the $f_i\lambda_i$  of {\Lyb}
and {\Lye} are quite different, with a ratio of about 11:1 between them, the
simultaneous agreement between the models and observations gives us additional
confidence in the measurement of $N_{\rm H\,I}$. A chi-square corresponding to
the best fit was denoted as $\chi_0^2$ for each component. Then, the upper and
lower limits of $N_{\rm H\,I}$ were calculated when the chi-squares were $\chi_0^2+1$.
The measured $N_{\rm H\,I}$ results are listed in Table \ref{tabNcol}. The Doppler
$b$ parameter of the {\HI} Lyman absorption lines for component $T1$ is estimated
to be $b_{\rm H\,I}=13.95\pm1.40\,{\rm km\,s^{-1}}$, which will be used to estimate
the gas temperature in Section \ref{secT} together with $b_{\rm Si\,IV}$ as shown
in Section \ref{secNcolSiIV}. The $b_{\rm H\,I}$ for the seven components are
listed in Table \ref{tabNcol}.

\begin{deluxetable*}{lccccccccc}
\tabletypesize{\scriptsize}
\tablewidth{0pt}
\tablecaption{Estimated Absorption-line Properties, Ionic Column Densities and Metallicities of the Absorber in SDSS J1220+0923}
\tablehead{
    \colhead {Parameters}     &
		\colhead {$T1$}           &
		\colhead {$T2$}           &
		\colhead {$T3$}           &
		\colhead {$T4$}           &
		\colhead {$T5$}           &
		\colhead {$T6$}           &
		\colhead {$T7$}           &
		\colhead {Total}
 				}
				\startdata
$z_{\rm abs\tablenotemark{a}}$  & $3.1411$ & $3.1424$  & $3.1447$ & $3.1461$ & $3.1479$ & $3.1491$ & $3.1497$ & $3.1447$ \\

$\log N_{\rm H\,I}\,({\rm cm^{-2}})$  & $14.75\pm0.05$ & $14.83\pm0.02$  & $15.37\pm0.02$ & $13.80\pm0.04$ & $14.00\pm0.01$ & $13.73\pm0.01$ & $13.66\pm0.01$ & $15.59\pm0.02$ \\

$b_{\rm H\,I}\,({\rm km\,s^{-1}})$  & $13.95\pm1.40$ & $19.05\pm1.90$   & $22.99\pm2.30$ &$28.17\pm3.00$  &  $23.64\pm2.13$  & $15.01\pm1.50$  & $25.35\pm2.03$  &  $\cdots$ \\
                                    &  $\cdots$& $19.00\pm1.90$   & $21.80\pm2.30$ &$105.64\pm3.00$  &      $19.94\pm1.79$  & $\cdots$ & $\cdots$ & $\cdots$ \\
                                                 & $\cdots$                        &  $\cdots$                & $10.56\pm2.30$ & $\cdots$ & $\cdots$ & $\cdots$ & $\cdots$ & $\cdots$ \\
$\log N_{\rm Si\,IV}\,({\rm cm^{-2}})$&$13.10\pm0.03$& $13.32\pm0.02$& $14.07\pm0.02$& $<12.240$       & $12.48\pm0.09$& $12.63\pm0.05$& $12.36\pm0.13$& $14.21\pm0.02$\\
$\log N_{\rm N\,V}\,({\rm cm^{-2}})$  &$13.13\pm0.04$& $14.43\pm0.02$& $14.98\pm0.02$& $13.85\pm0.02$& $14.47\pm0.01$& $13.55\pm0.01$& $13.87\pm0.01$& $15.24\pm0.02$\\

$\log N_{\rm C\,III}\,({\rm cm^{-2}})$  &$13.51\pm0.04$& $14.05\pm0.08$& $>14.95$& $12.89\pm0.05$& $13.29\pm0.03$& $13.28\pm0.03$& $13.30\pm0.05$& $>15.04$\\
$\log N_{\rm C\,IV}\,({\rm cm^{-2}})$  &$<13.63$& $>14.67$& $>15.61$& $13.91\pm0.04$& $14.47\pm0.03$& $13.98\pm0.03$& $14.21\pm0.03$& $>15.72$\\
$\log N_{\rm N\,III}\,({\rm cm^{-2}})$&$13.47\pm0.06$ &$<13.70$ & $\cdots$& $\cdots$& $<12.88$& $<13.02$& $<12.76$&$\cdots$ \\
$\log N_{\rm N\,III*}\,({\rm cm^{-2}})$&$<12.82$ & $<13.04$& $<13.79$&$12.64$ &$<12.20$ & $<12.35 $&$<12.08$ & $<13.95$ \\
{$Z\,(Z_{\sun})$}  &$1.50\pm0.25$ & $\cdots$& $\cdots$ & $\cdots$& $\cdots$ & $\cdots$ & $\cdots$ & $\sim3.25$ \\
{$Z_{\rm min}\,(Z_{\sun})$}  &$1.50$ & $2.08$ & $3.08$ & $2.12$ & $2.25$ & $5.76 $ & $3.77$ & $2.49$

				\enddata
				\tablenotetext{}{Notes. The column densities are reported in logarithmic terms with units of $\rm cm^{-2}$, while the metallicities are presented in their actual values relative to the solar abundance.}
				\tablenotetext{a}{Redshifts of the absorption lines, calculated from {\NV} absorption lines, are weighted by optical depths for individual components and the absorptions as a whole.}
				\label{tabNcol}
				\end{deluxetable*}

\subsubsection{{Ground and Excited States of {\NIII}}}
\begin{figure*}[htb]
\center{\includegraphics[width=19cm]  {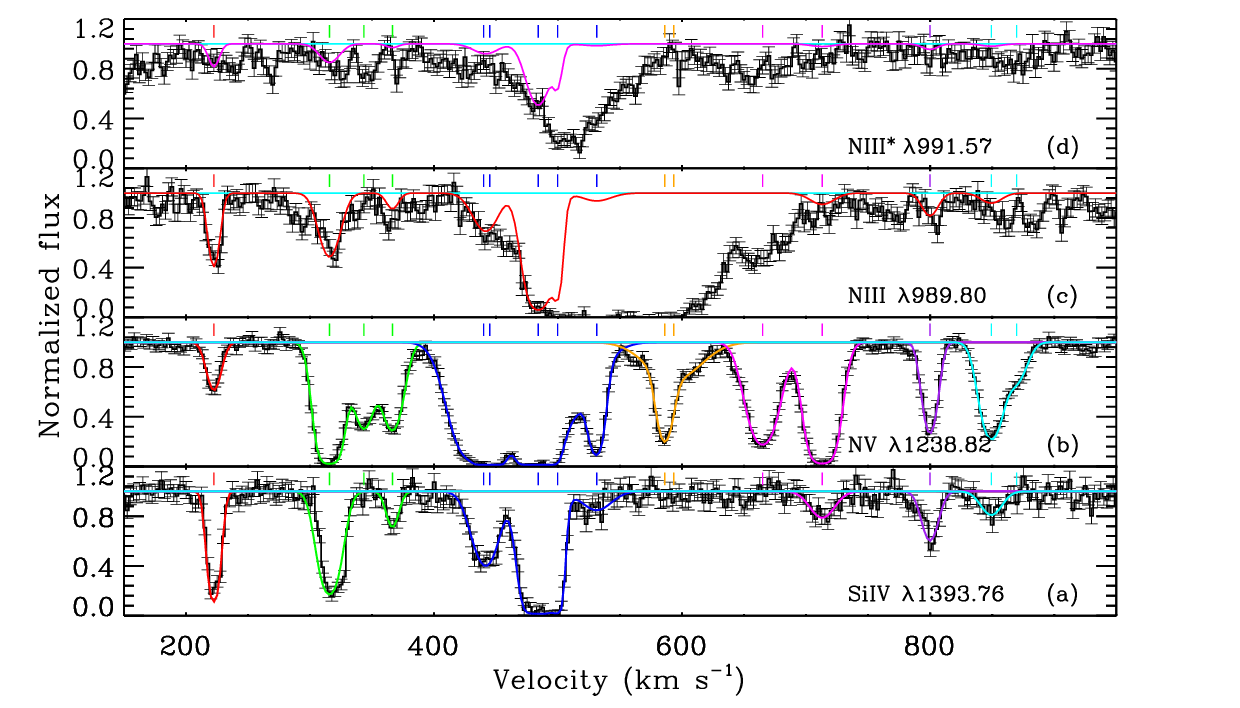}}
\caption{The normalized spectra in the \ion{N}{3} and \ion{N}{3*} regions with
respect to \ion{N}{3} $\lambda989.80$ and \ion{N}{3*} $\lambda991.57$. The
velocity structures of {\SiIV} $\lambda1393.76$ and {\NV} $\lambda1238.82$ are also
replotted in panels (a) and (b) for comparison. The small vertical lines in each panel
indicate the centers of the fitted components.
}
\label{figNIII}
\end{figure*}

We show the velocity structures of  {\NIII} $\lambda989.80$ and {\NIII*} $\lambda991.57$,
the ground and excited states for the N$^{2+}$ ion, in Figure \ref{figNIII}. We also show
the velocity structures of {\SiIV} $\lambda1393.76$ and {\NV} $\lambda1238.82$,
replotted from Figure \ref{figSiIVNV}, in Figures \ref{figNIII}(a) and (b), for
comparison. The {\NIII} wavelength region shows clear absorptions at the velocities
corresponding to those of {\SiIV} and {\NV}. The absorption component $T1$ is
free from contamination; however, the other absorption components suffer from
more or less contamination. We replotted the fitted {\SiIV} $\lambda1393.76$
total absorption profile to the {\NIII} and {\NIII}$^*$ regions, after applying
a scale factor to match the absorption troughs associated with component $T1$
of {\NIII} and {\NIII}$^*$, as shown by the red and magenta lines in Figures
\ref{figNIII}(c) and (d). Assuming that {\NIII} absorbing gas has the
same covering fraction as those of {\SiIV} and {\NV}, we use one Voigt function
to fit the absorption trough of component $T1$, and the fitted result
agrees well with the observation. The obtained $N_{\rm N\,III}$ of component
$T1$ is listed in Table \ref{tabNcol}. By shifting the modeled {\SiIV} $\lambda1393.76$
absorption profiles for components $T2$, $T5$, $T6$, and $T7$ to the {\NIII} wavelength
and using them as templates to match the {\NIII} region, we estimated the upper
limit of $N_{\rm N\,III}$ for these components (except for the heavily contaminated
components $T3$ and $T4$), and the results are listed in Table \ref{tabNcol}.

Different from the {\NIII} $\lambda989.80$, {\NIII*} $\lambda991.57$ does not have
clear troughs at the velocity where the {\SiIV} and {\NV} show absorption troughs. By shifting
the absorption profile of {\SiIV} to the {\NIII*} velocity and using it as a template
to match the {\NIII*} region, we estimated the upper limit of $N_{\rm N\,III}*$, as
listed in Table \ref{tabNcol}. A clear detection of {\NIII} $\lambda989.80$ in
combination with an almost nondetection of {\NIII*} $\lambda991.51$ suggests
that the absorbing gas is of low density (see Section \ref{secnH}).

\subsection{Gas Temperature}\label{secT}
The Doppler $b$ parameter, contributed by both thermal and turbulent motions
in gas, is described as \citep{Finn2014}
\begin{equation}
b^2=b_0^2+\frac{2kT}{m},
\end{equation}
where $b_0$ represents the turbulent contribution to the line width, $k$ is
the Boltzmann constant, $T$ is the gas temperature, and $m$ is the atomic mass
of the ions in question. Since {\HI} and {\SiIV} show a similar velocity
structure (see Figure \ref{fig:HI}), we hypothesize that {\HI} and {\SiIV} absorptions
arise from the same gas, implying that they have the same temperature and $b_0$ value. 
Consequently, the gas temperature can be derived as
\begin{equation}
T=\frac{b_{\rm H\,I}^2-b_{\rm Si\,IV}^2}{2k(\frac{1}{m_{\rm H\,I}}-\frac{1}{m_{\rm Si\,IV}})}.
\end{equation}
Using the $b$ values for {\HI} and {\SiIV} of the detached absorption
component $T1$ estimated in Section \ref{secNcolHI} and \ref{secNcolSiIV},
the gas temperature for this component can be estimated as
$T=(9.99\pm2.76)\times10^3\,\rm K$.

\section{Physical Properties of the Absorber} \label{secphys}
\subsection{Hydrogen Number Density constraint}\label{secnH}
We employ the photoionization model generated with \texttt{Cloudy} version c17
\citep{Ferland1998,Ferland2017}, in conjunction with the
observed ionic column densities, to constrain the physical properties of SDSS
J1220+0923. A slab-shaped dust-free model for absorbing gas with homogeneous
density and scaled solar abundance is used to regenerate the measured ionic
column densities. We applied the incident spectral energy
distribution (SED), characterized primarily by a blackbody big
bump (hereafter denoted as 4B SED), which is
a superposition of a power-law UV bump component and a power-law X-ray component
described as $\nu^{\alpha_{\rm UV}}{\rm exp}(-h\nu/kT_{\rm BB})\,{\rm exp}(-kT_{\rm IR}/h\nu)$
and $a\nu^{\alpha_{\rm X}}$, respectively. They are extracted and combined from
multiband observations \citep{Zamorani1981,Elvis1994}, and are considered typical
for quasars \citep{Ferland2017}. The UV bump is parameterized by a UV power-law
index $\alpha_{\rm UV}=-0.5$ \citep{Elvis1994}, and is exponentially cut off with
temperature $T_{\rm BB}=1.5\times10^5\,\rm K$ at the high-energy end and
$T_{\rm IR}=1580\,{\rm K}$ at the low-energy end. The power-law X-ray component
has an index $\alpha_{\rm X}=-2$ greater than 100 keV and $-1$ between 13.6\,\rm keV
and 100 keV \citep{Elvis1994}. The flux ratio of the X-ray to optical is
$\alpha_{\rm OX}=-1.4$ \citep{Zamorani1981}.

There are four parameters that are critical in characterizing the physical
conditions of the absorbing gas: $Z$, $U$, {\nH} and {\NH}. The other properties,
such as the distance ($r_{\rm abs}$) of the gas to the central BH, the
mass-flow rate ($\dot{M}$), and the kinetic luminosity ($\dot{E}_{\rm k}$), can
be derived from these four parameters \citep[e.g.,][]{Borguet2012b}. However,
it is a challenge to find the best model that can simultaneously reproduce
the observed column densities of multiple ions in a four-dimensional parameter
space ($U$, {\nH}, {\NH}, and $Z$). Thus, we seek additional constraints on these
four parameters to reduce the number of free parameters needed for the modeling.

The column density ratio between excited and ground states for the same ion depends
 on the electron density ($n_{\rm e}\approx1.2n_{\rm H}$) and temperature \citep[e.g.,][]{Osterbrock2006}.
The ion ratios, such as those for {\CII}, {\SiII} \citep[e.g.,][]{ Borguet2012b},
{\NIII}, {\SIII} \citep[e.g.,][]{Chamberlain2015}, and {\SIV} \citep[e.g.,][]{Xu2018, Xu2019},
are often used to estimate the {\nH} by assuming a typical temperature, for instance,
$T\simeq10,000\,\rm K$, for plasma photoionized by a quasar's spectrum \citep[e.g.,][]{Arav2018}.
In this work, we use the ratio of $N_{\rm N\,III^*}/N_{\rm N\,III}$ to establish the
relationship between {\nH} and ($U$, $Z$) under the condition when the simulated
$N_{\rm H\,I}$ matches the observed value.

\begin{figure}[htb]
\center{\includegraphics[width=8.4cm]  {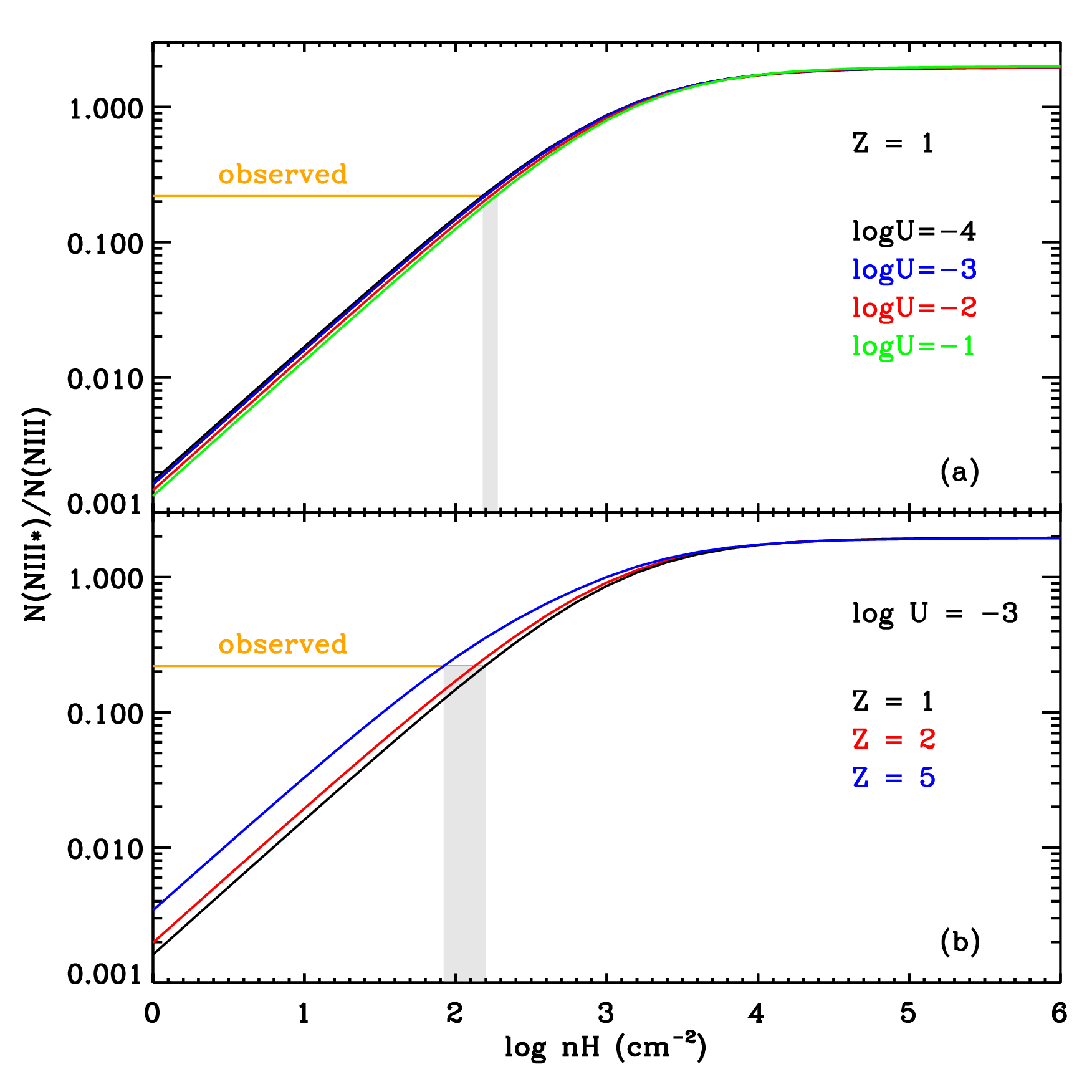}}
\caption{The simulated $N_{\rm N\,III^*}/N_{\rm N\,III}$ as a function of
{\nH}, at the condition when the simulated $N_{\rm H\,I}$ reaches the
observed value of component $T1$, for (a) $Z=Z_\sun$ with different $U$,
and (b) ${\rm log}\,U=-3$ with different $Z$. The observed value of
$N_{\rm N\,III^*}/N_{\rm N\,III}$ is indicated as the orange straight lines.
}
\label{figratio}
\end{figure}

\begin{figure}[htb]
\center{\includegraphics[width=8.4cm]  {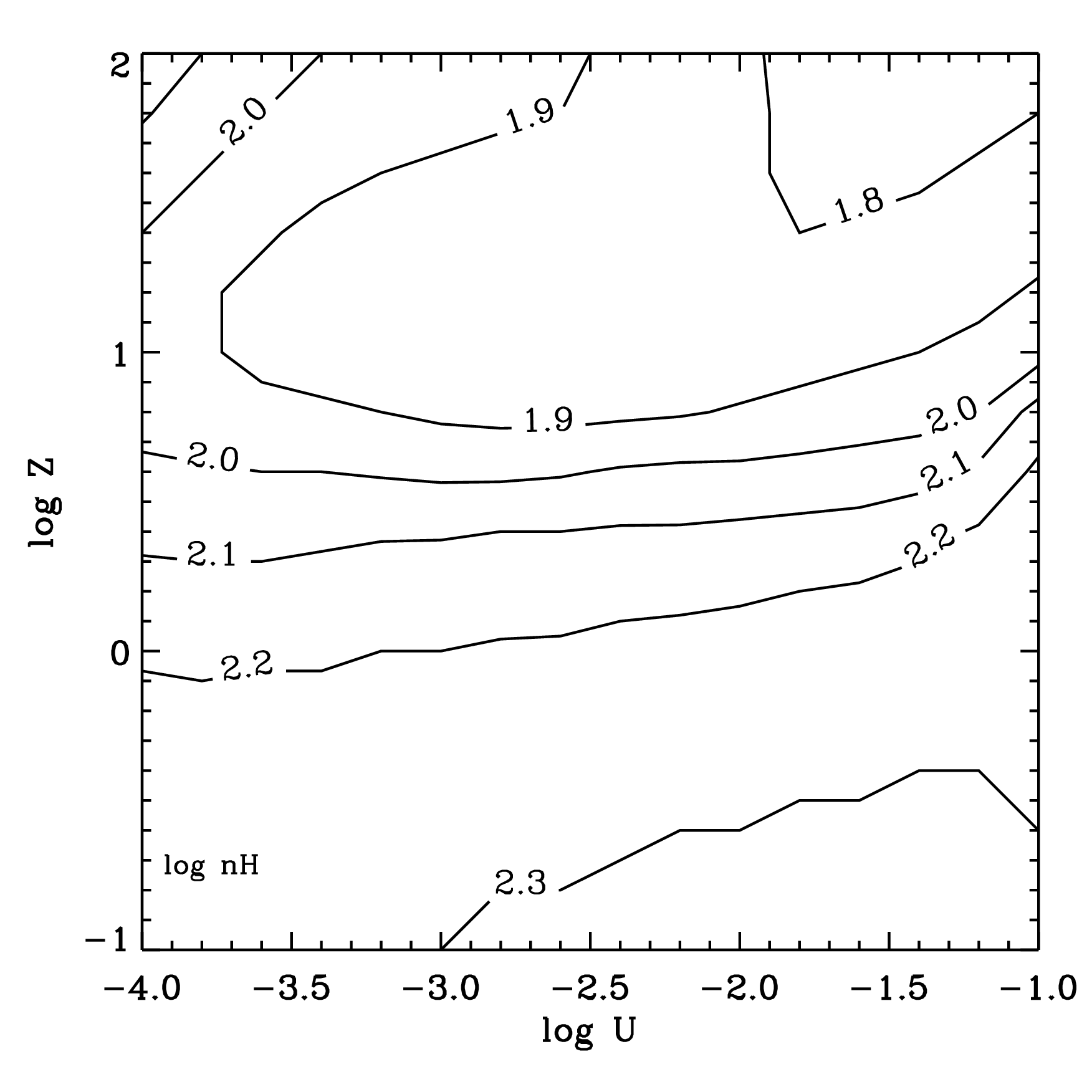}}
\caption{The {\nH} as functions of $U$ and $Z$ for component $T1$
obtained by the $N_{\rm N\,III}*/N_{\rm N\,III}$ when the simulated $N_{\rm H\,I}$
reaches the observed value (see the text for the details). The corresponding {\NH} contours
are plotted in Figure \ref{figNHUZ}.
}
\label{fignHUZ}
\end{figure}

\begin{figure}[htb]
\center{\includegraphics[width=8.4cm]  {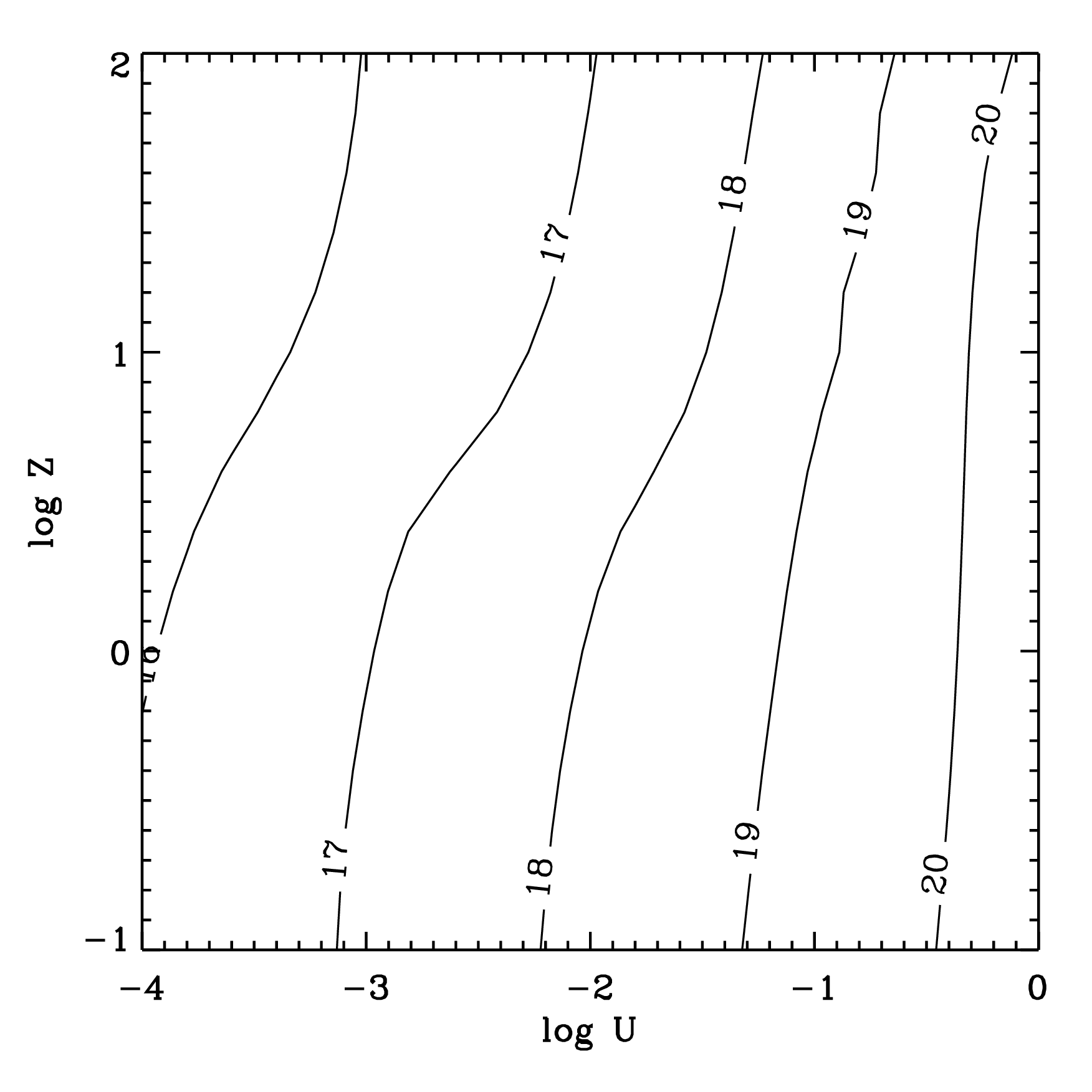}}
\caption{The {\NH} as a function of $U$ and $Z$ in the condition when the
modeled $N_{\rm H\,I}$ reaches the observed value of component $T1$.
The {\nH} values used to obtain the {\NH} contours are functions of
$U$ and $Z$, as shown in Figure \ref{fignHUZ}.
}
\label{figNHUZ}
\end{figure}

For a given ($U$ and $Z$), the $N_{\rm N\,III^*}/N_{\rm N\,III}$ can be calculated
as a function of {\nH} at the condition when the simulated $N_{\rm H\,I}$ at a given
parameter set of ($U$, $Z$, $n_{\rm H}(i)$) reaches the observed value
(taking ${\rm log}\,N_{\rm H\,I}=14.754\,\rm cm^{-2}$ for component $T1$ as an
example). We plot the curves between $N_{\rm N\,III^*}/N_{\rm N\,III}$ and {\nH}
for solar abundance with ${\rm log}\,U=$ $-4$, $-3$, $-2$, and $-1$ in Figure
\ref{figratio} (a), and for ${\rm log}\,U=-3$ with $Z=1$, 2, and 5 in Figure
\ref{figratio} (b). It can be seen that the curves are almost identical for
different $U$ at a fixed $Z$, while there are only slight differences when $Z$
is different at a fixed $U$. This means that in the range of
$1\lesssim{\rm log}\,n_{\rm H}\,(\rm cm^{-3})\lesssim5$, the {\nH} can be
determined from the ratio of $N_{\rm N\,III^*}$ to $N_{\rm N\,III}$. Thus,
the {\nH} for each given ($U$, $Z$) can be derived by the observed ratio
between $N_{\rm N\,III^*}$ and $N_{\rm N\,III}$. Then, the simulations were
conducted over $-4\leqslant{\rm log}\,U\leqslant-1$ and $-1\leqslant{\rm log}\,Z\leqslant2$,
with a step of $0.2\rm\,dex$. Consequently, the relations between {\nH} and
($U$ and $Z$) can be built as
\begin{equation}\label{eqnUZ}
n_{\rm H}=f(U,\,Z).
\end{equation}
We plot the {\nH} and {\NH} contours, taking component $T1$ as an example, as a
function of $U$ and $Z$ in Figures \ref{fignHUZ} and \ref{figNHUZ}, respectively.

Therefore, with the help of the observed $N_{\rm H\,I}$ and the ratio
of $N_{\rm N\,III}*$ to $N_{\rm N\,III}$, we are able to decrease
the number of parameters for the model fitting from four ($U$, {\nH},
{\NH}, and $Z$) to two ($U$ and $Z$). The best-fitting model
that can simultaneously reproduce the observed column densities
of detected ions can now be explored in a 2D parameter space
($U$ and $Z$).

\subsection{Results of the Best-fit Models} \label{secUnN}

\begin{figure}[htb]
\center{\includegraphics[width=8.4cm]  {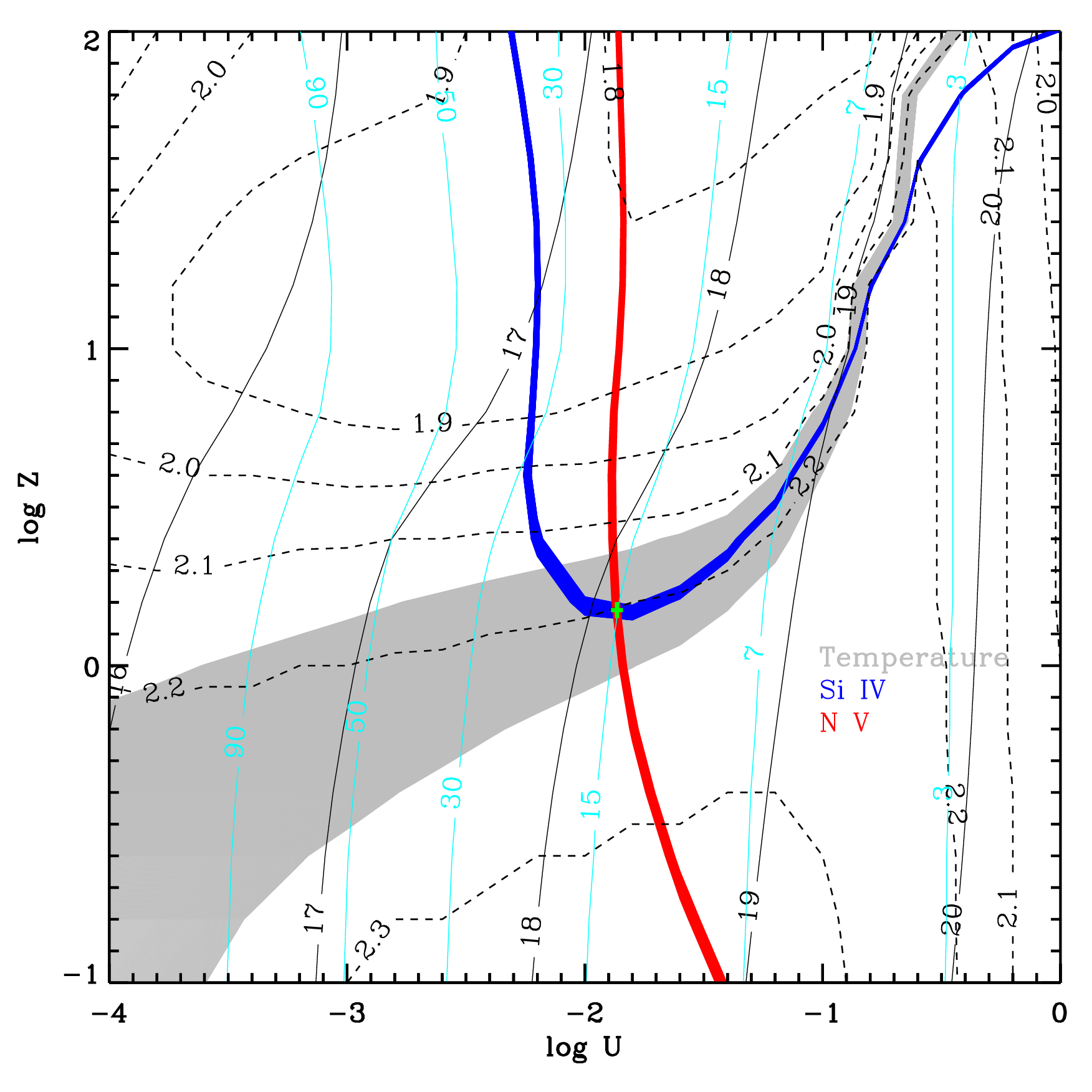}}
\caption{Photoionization models for component $T1$ in SDSS J1220+0923. The blue,
red, and gray shaded regions represent the parameter space in ($U$, $Z$, {\nH},
and {\NH}) allowed by the observed $N_{\rm Si\,IV}$ and $N_{\rm N\,V}$ and
temperature, within $1\sigma$ error. The black solid and dashed lines represent
the {\nH} and {\NH} as functions of $U$ and $Z$ (see Figures \ref{fignHUZ} and
\ref{figNHUZ}). The cyan lines represent the location of the absorber
($r_{\rm abs}$) as functions of $U$ and $Z$. The best model is marked by the
green cross. The errors of $U$, $Z$, {\nH}, and {\NH} are approximately estimated
by the difference between the edge values of the intersection region and the
best model.
}
\label{figmodT1}
\end{figure}

First, we estimated the $U$, $Z$, {\nH} and {\NH} for component $T1$.
For a set value of ($U$ and $Z$), {\nH} and {\NH} can be determined via the {\HI}
column density and the column density ratio between the excited and ground states
of {\NIII} (Figures \ref{fignHUZ} and \ref{figNHUZ}). With these four values,
the absorption strengths of various metal ions were simulated, which can be
compared with the observed absorption lines. In Figure \ref{figmodT1}, we show
the parameter spaces in ($U$ and $Z$) that are permitted and excluded by the
observed column densities of different ions. The parameter spaces suggested by
the measured column densities of {\SiIV} and {\NV} are shown as the blue and red
regions, respectively. The best model is the intersection of these two areas,
indicated by the green cross. It gives the best-estimated model of
${\rm log}\,U=-1.865\pm0.025$, ${\rm log}\,Z=0.175\pm0.025$,
${\rm log}\,n_{\rm H}=2.203\pm0.008\,{\rm cm^{-3}}$, and
${\rm log}\,N_{\rm H}=18.12\pm0.03\,{\rm cm^{-2}}$. The parameter space
suggested by the measured temperature is shown as the gray region,\footnote{For
our source, the gas is relatively thin, resulting in minimal temperature
variation inside. For example, the simulated temperature variation using
the best model of (${\rm log}\,U=-1.865$, ${\rm log}\,Z=0.175$,
${\rm log}\,n_{\rm H}=2.203\,\rm cm^{-3}$, and
${\rm log}\,N_{\rm H}=\rm 18.12\,\rm cm^{-2}$) is only $241\,\rm K$. Consequently,
we selected the temperature at the outermost layer, corresponding to the
{\NH} where the simulated $N_{\rm H\,I}$ reaches the measured value.} which
covers the best model. The temperature, as an independent validation,
provides us with additional confidence in the best model.

In order to calculate the errors of the four parameters propagated from the error
of $N_{\rm H\,I}$, two figures similar to Figure \ref{figmodT1} are drawn, using
the upper and lower limits of $N_{\rm H\,I}$ as listed in Table \ref{tabNcol},
to derive the corresponding values of the four parameters, which are treated as
the lower and upper limits. The errors of {\nH} and $U$, contributed from the
error of $N_{\rm H\,I}$, are negligible and thus ignored, while the errors of
$Z$ and {\NH}, resulting from the error of $N_{\rm H\,I}$, are determined to
be 0.05 and $0.060\,\rm dex$, respectively. The values of $Z$ and {\NH}, including
the propagated errors, are updated as $\log Z=0.175\pm0.075$, and
$\log N_{\rm H}=18.12\pm0.09\,\rm cm^{-2}$. The predicted column densities
of {\NIII}, {\SiIII}, and {\SiII} for component $T1$ by the best model in
Figure \ref{figmodT1} are ${\rm log}\,N_{\rm N\,III}=13.77\pm0.03\,\rm cm^{-2}$,
${\rm log}\,N_{\rm Si\,III}=12.68\pm0.04\,\rm cm^{-2}$, and ${\rm log}\,N_{\rm Si\,II}=11.18\pm0.08\,\rm cm^{-2}$.
This model also predicts ${\rm log}\,N_{\rm C\,III}=14.34\pm0.03\rm\,cm^{-2}$
and ${\rm log}\,N_{\rm C\,IV}=14.20\pm0.04\rm\,cm^{-2}$ for component $T1$.
The modeled $N_{\rm N\,III}$ is larger than that of the measured result (see Table
\ref{tabNcol}) by $0.3\,\rm dex$, and the modeled $N_{\rm C\,III}$ and $N_{\rm C\,IV}$
are larger than those of the measured results (see Table \ref{tabNcol}) by 0.8 and $0.6\rm\,dex$,
which will be discussed in Section \ref{secnonsolar}.


With the measured {\nH} and $U$ of the absorbing gas, the distance ($r_{\rm abs}$)
of the absorber away from the central source can be estimated. The definition of
$U$ is
\begin{equation}\label{eqU}
U=\frac{Q_{\rm H}}{4\pi r_{\rm abs}^2 n_{\rm H}c},
\end{equation}
where $Q_{\rm H}$ is the total rate of hydrogen-ionizing photons emitted by the
central source \citep[e.g.,][]{Dunn2010a}. $Q_{\rm H}$ equals to
$L(<912)/\overline{E_{\rm ph}(<912)}$, where $L(<912)$ is the ionizing luminosity
of the continuum source, and  $\overline{E_{\rm ph}(<912)}$ is the average energy
for all ionizing photons. $Q_{\rm H}=1.73\times10^{57}\,{\rm s^{-1}}$, which is
estimated by matching the flux of the model SED to the dereddened observed flux
at 1450 {\AA} in the SDSS rest-frame spectrum using a standard cosmology
($H_0=73.0\,{\rm km\,s^{-1}\,Mpc^{-1}}$, $\Omega_\Lambda=0.73$, and $\Omega_m=0.27$).
The distance of component $T1$ was determined to be $14.8\pm0.5\, \rm kpc$
for our best model.

We also experimented with the MF87 SED \citep{Mathews1987} as an input to our
models for component $T1$. The main difference between this SED and 4B SED
is the location of their peaks, which are at approximately 3.0 and 0.5 Rydbergs,
respectively. The best model results from the MF87 SED are
${\rm log}\,Z=0.36\pm0.04$, ${\rm log}\,U=-1.95\pm0.015$,
${\rm log} \,N_{\rm H}=17.7\pm0.05$, ${\rm log}\,n_{\rm H}=2.1\pm0.02$, and $r_{\rm abs}=31\pm1\,\rm kpc$.
The $U$ and {\nH} are only slightly smaller than those estimated from the 4B SED
by 0.09 and $0.1\rm\,dex$, and {\NH} is smaller than that from the 4B SED by $0.4\,\rm dex$.
The MF87 SED suggests a higher metallicity and further distance, with $Z$ and
$r_{\rm abs}$ values about twice the values in models using the 4B SED.
In this work, we choose to use the model with the 4B SED to present our results.
It can be seen that component $T1$ of the inflowing gas, located at $\sim15\,\rm kpc$
from the central BH, exhibits supersolar abundance even with the more conservative
model using the 4B SED.

\begin{figure}[htb]
\center{\includegraphics[width=8.4cm]  {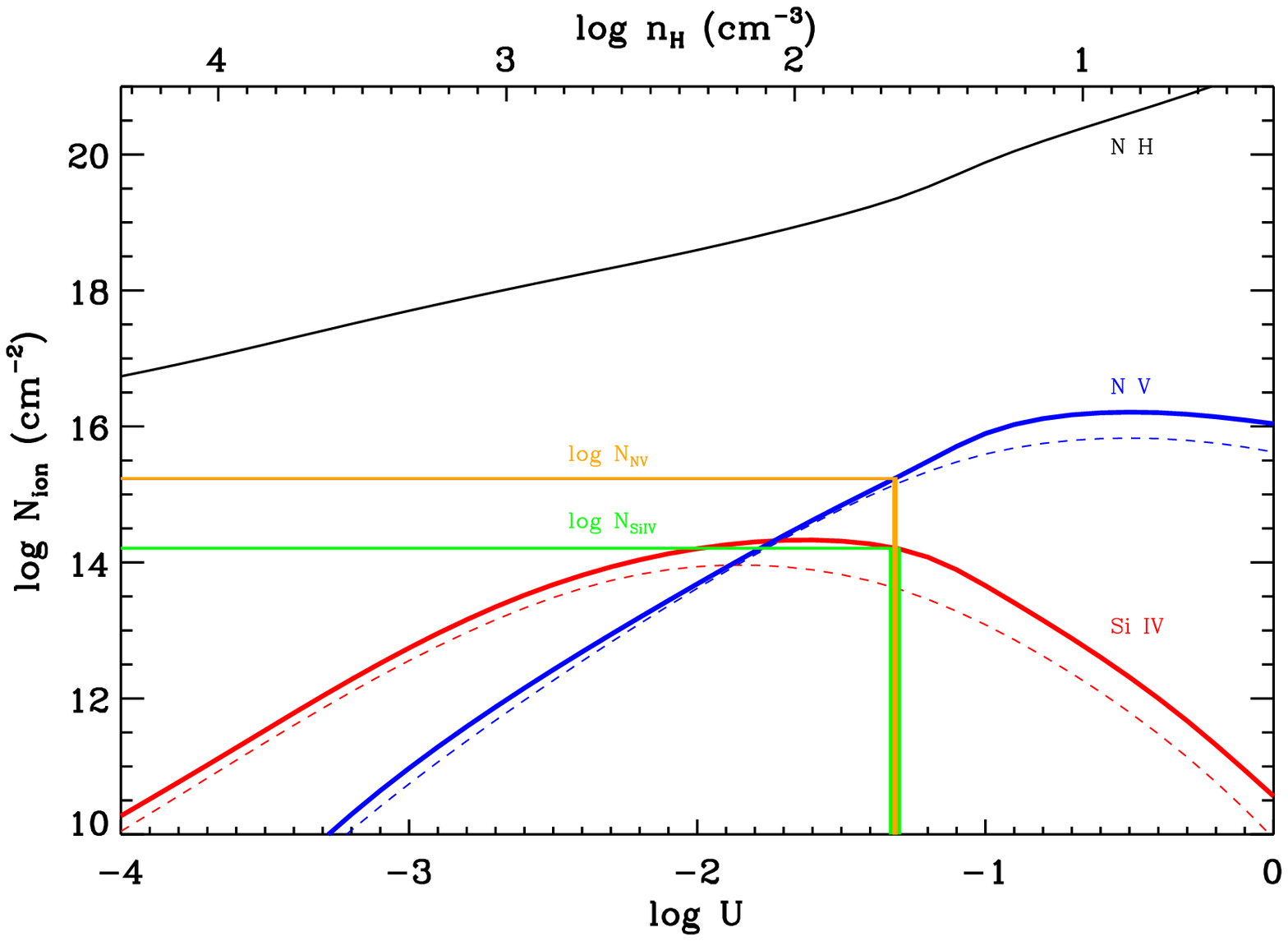}}
\caption{Photoionization models of the absorbing gas as a whole in SDSS J1220+0923
assuming the radial distance ($r_{\rm abs}$) to central BH is identical to
that of component $T1$ (see the text for the details).
}
\label{figmodtotal}
\end{figure}

Second, we probe the nature of the absorber as a whole by assuming that it has
an identical distance from the central BH as that of component $T1$. This
assumption, upon inspection
of Equation (\ref{eqU}), suggests a relation of $n_{\rm H}=2.18/U$. As an attempt,
assuming that the  abundance of the absorber as a whole is the same as that of
component $T1$, the total $N_{\rm Si\,IV}$ and total $N_{\rm N\,V}$ were calculated at the
condition when the simulated $N_{\rm H\,I}$ matches the observed value. They
are plotted as functions of $U$ (also {\nH}) in Figure \ref{figmodtotal},
as shown by the dashed blue and red lines. It can be seen that the simulated
total $N_{\rm Si\,IV}$ over the range of $-4\leqslant{\rm log}\,U\leqslant0$ cannot
reach the observed value, as shown by the horizontal green line. We then tried
different metallicities near that of component $T1$ to search for the best $U$
(and also {\nH}) that could simultaneously reproduce the observed total $N_{\rm Si\,IV}$
and total $N_{\rm N\,V}$. It is found that the simulated $N_{\rm Si\,IV}$ and $N_{\rm N\,V}$
increased with increasing metallicity. The solid red and blue curves in Figure
\ref{figmodtotal} show the simulated $N_{\rm Si\,IV}$ and $N_{\rm N\,V}$ for
$Z=3.54\,Z_\sun$. The best model leads to ${\rm log}\,U=-1.31\pm0.02$, as
indicated by the vertical orange and green lines, which can simultaneously
reproduce the observed total $N_{\rm Si\,IV}$ and total $N_{\rm N\,V}$. The corresponding
hydrogen number density is ${\rm log}\,n_{\rm H}=1.66\pm0.03\,{\rm cm^{-3}}$.
The {\NH} is shown by the black line in Figure \ref{figmodtotal}, and the best
model is ${\rm log}\,N_{\rm H}=19.3\pm0.02\,{\rm cm^{-2}}$. Furthermore,
the $N_{\rm C\,III}$ and $N_{\rm C\,IV}$ predicted by the best model are
${\rm log}\,N_{\rm C\,III}=15.58\,\rm cm^{-2}$ and ${\rm log}\,N_{\rm C\,IV}=15.78\,\rm cm^{-2}$,
which are consistent with the measured lower limits of $N_{\rm C\,III}$ and $N_{\rm C\,IV}$ as shown
in Table \ref{tabNcol}, serving as additional confirmation of the model.
This characteristic $Z$ value suggests that the inflowing gas in our source could be metal-enhanced.

It can be seen in Figure \ref{figSiIVNV} that the relative strengths of
the {\SiIV} and {\NV} absorption lines for component $T1$ are different from
the other components. The $N_{\rm Si\,IV}$ and $N_{\rm N\,V}$ for
component $T1$ are almost identical (see Table \ref{tabNcol}), while for
all the other components, $N_{\rm Si\,IV}$ is lower than $N_{\rm N\,V}$
by about $1\rm\,dex$. Since the ionization energy of $\rm N^{3+}$ is $77.47\,\rm eV$
(the energy required to create $\rm N^{4+}$ from $\rm N^{3+}$), and is
significantly larger than that of $\rm Si^{2+}$ at $33.49\,\rm eV$, these
differences suggest that the component $T1$ might have a smaller $U$ than the other
components. This further suggests that the component $T1$ has a larger {\nH},
see Equation (\ref{eqU}), assuming the different components are located at
the same distance from the central BH. Indeed, this is very likely
true since the {\nH} of the absorber as a whole is $\sim10^{1.66}\,\rm cm^{-3}$,
less than that of component $T1$ at
$\sim10^{2.20}\,\rm cm^{-3}$.

\section{Discussion} \label{secdiss}

\subsection{The Nature and Origin of the Absorber}
Through the detailed modeling, we have shown that the absorbing gas of SDSS J1220+0923
is at $\sim15\,\rm kpc$ from the central BH. This measurement is
consistent with the absorbing gas having a low-density {\nH} $\sim 10^{2}\,\rm cm^{-3}$
(see Section \ref{secUnN}). The {\nH} is expected to be low for gas at larger distances
from the central BH. Indeed, numerous studies have reported the properties of
absorbers located at several or dozens of kiloparsecs, similar to SDSS J1220+0923
(e.g., $\sim28\,\rm kpc$, \citealp{Hamann2001}; $\sim3.3\,\rm kpc$, \citealp{Moe2009};
$\sim6$ and $\sim17\,\rm kpc$, \citealp{Dunn2010b}; $\sim3.7\,\rm kpc$, \citealp{Aoki2011};
$>3\,\rm kpc$, \citealp{Edmonds2011}; $\sim10\,\rm kpc$, \citealp{Borguet2012b};
$\sim3.4\,\rm kpc$, \citealp{Arav2013}; $3.3-6\,\rm kpc$, \citealp{Finn2014}).
The {\nH}  for these absorbers ranges from $10^{1.4}$ \citep{Edmonds2011} to
$10^{3.75}\,\rm cm^{-3}$ \citep{Moe2009}. They are considerably small compared
with the absorbers located at parsec or dozens of parsecs distance from the central
source \citep[e.g.,][]{Zhang2015, Liu2016, Shi2016, Veilleux2016, Tian2021}. For example,
the outflowing gas at $1-30\rm \,pc$ from the central source is reported to have
$7.5<{\rm log}\,n_{\rm H}<9.5$ \citep{Kool2002}.

Another intriguing feature of the absorbing gas of SDSS J1220+0923 is that it is
red-shifted relative to the systematic redshift of the quasar, suggesting it is
an inflowing gas. Assuming that the absorbing gas can be described by a simple
picture of a partially filled shell, the mass inflow rate can be estimated as
\citep{Borguet2012b},
\begin{equation}\label{eqM}
\dot{M}_{\rm infow}=4\pi r_{\rm abs}\Omega\mu m_{\rm p}N_{\rm H}v,
\end{equation}
where $\mu=1.4$ is the mean relative atomic mass per proton, $m_{\rm p}$ is the mass of a
proton, and $\Omega$ is the global covering factor of the inflowing gas clouds. The
$r_{\rm abs}$ and {\NH} of the absorber as a whole have been obtained, as shown
in Section \ref{secUnN}. The $\tau(v)$-weighted mean velocity, $v$, using Equation
(\ref{eqtau}), calculated by the {\SiIV} $\lambda1402.77$ absorption lines is
$\sim470\,\rm km\,s^{-1}$. Then, we have $\dot{M}_{\rm inflow}\sim20.0\,\Omega\,M_\sun\,\rm yr^{-1}$.
We roughly estimate the $\Omega$ using the detected fraction of NAL quasars with
associated {\NV} lines, which is $\sim33\%$ reported by S. Perrotta et al. (\citeyear{Perrotta2016}).
Then, the $\dot{M}_{\rm inflow}$ is estimated to be $\sim6.6\,M_\sun\,\rm yr^{-1}$.

Outflowing gases are generally metal-rich, suggested not only from simulations 
\citep[e.g.,][]{Christensen2018,Vijayan2024}, but also observations 
\citep[e.g.,][]{Turner2015,Chisholm2018}. In fact, solar or supersolar abundances 
in high-redshift Quasars have been reported across regions from their broad-line 
regions (BLRs) to those that are a few hundred kiloparsecs away (e.g., V. D'Odorico et al. 
\citeyear{DOdorico2004a}, \citeyear{DOdorico2004b}; S. Lai et al. \citeyear{Lai2022}), 
suggesting that the metal enrichment by the outflow can be very efficient. There 
is direct evidence as well; for example, the outflows in UM 680, at a large 
distance of $70-120\,\rm kpc$, have carbon and nitrogen abundances at least five 
times the solar value (V. D'Odorico et al. \citeyear{DOdorico2004a}, \citeyear{DOdorico2004b}).
In contrast, the inflowing gas originating from primordial IGM is metal-poor 
\citep[e.g.,][]{Dayal2013}, and the metal abundance of inflows in Markarian 
1486 is observed to be below 5\% $Z_\sun$ \citep{Cameron2021}. What is puzzling 
is where the metal-strong inflowing gas in our source, at $\sim15\,\rm kpc$, 
comes from. There are some possible origins of the gas at 15 kpc scale, including 
(1) IGM, (2) a companion galaxy, (3) returning outflows, and (4) inflowing galactic 
winds. The first two origins are external, while the last two are internal.

Our absorber has a high metal abundance, making the first possible origin (the absorber
coming from the IGM) quite unlikely. We checked the optical and NIR images of 
SDSS J1220+0923 from the SDSS and UKIDSS observations, and no companion galaxy was 
detected. However, the second potential origin cannot be ruled out, as the companion 
galaxy may be too faint to be detected, especially considering the relatively high 
systemic redshift. Additionally, the limited spatial resolution of the available 
images may not be able to resolve the companion galaxies.

The large distance, low-density, relatively low-velocity
absorbing gas found in SDSS J1220+0923 is similar to that observed in
IRAS F224546-5125 \citep{Borguet2012b}. However, the absorption lines in their
source are blue-shifted, and they reported that their absorber exhibits
characteristics of an outflowing galactic wind. Is it possible that the
absorber in SDSS J1220+0923 is associated with an outflowing galactic wind, in
that the outflow gas decelerates as it goes out and eventually falls back?
The feasibility of this scenario can be considered together with the inflow
velocity of the absorbing gas. For a returning outflow gas, the inflow velocity
should be roughly consistent with the kinematics of the galaxy, as it is gravitationally
bound to the galaxy. Typically, the rotation velocities of massive galaxies at $\sim 15$ kpc
from their centers are $\lesssim 300\,\rm km\,s^{-1}$\citep{Bosma1978, Faber1979, Persic1995, Bell2001}.
While the  observed velocity of component $T1$ at $\sim220\,\rm km\,s^{-1}$ can be
consistent with the scenario, the high inflow velocities of other components,
for example, $\sim900\,\rm km\,s^{-1}$ for component $T7$, are hard to accommodate
in the picture. Even though our estimation of the distances has some uncertainties,
and we cannot firmly exclude the `returning outflow' scenario, we think this origin
of absorbing gas is unlikely.

Meanwhile, if the absorbing gas that we see simply comes from one episode of a
stellar wind and/or a supernova explosion, and if the gas is accidentally pushed
in the direction that is away from us, we would naturally see it as an inflowing
gas in our line of sight to the quasar. Interestingly, the high-abundance
feature of our absorbing gas can be easily explained in the scenario, as the gas may
contain heavy elements that are newly formed in the star. Stellar winds and/or
supernova explosions typically have velocities at hundreds of kilometers per second \citep[e.g.,][]{Johnstone2015},
which is also consistent with the inflow velocities that we measured for the absorbing gas.

\subsection{Diagnosis of High-metallicity Gas via the Absorption Lines}
\begin{figure*}[htb]
\center{\includegraphics[width=18cm]  {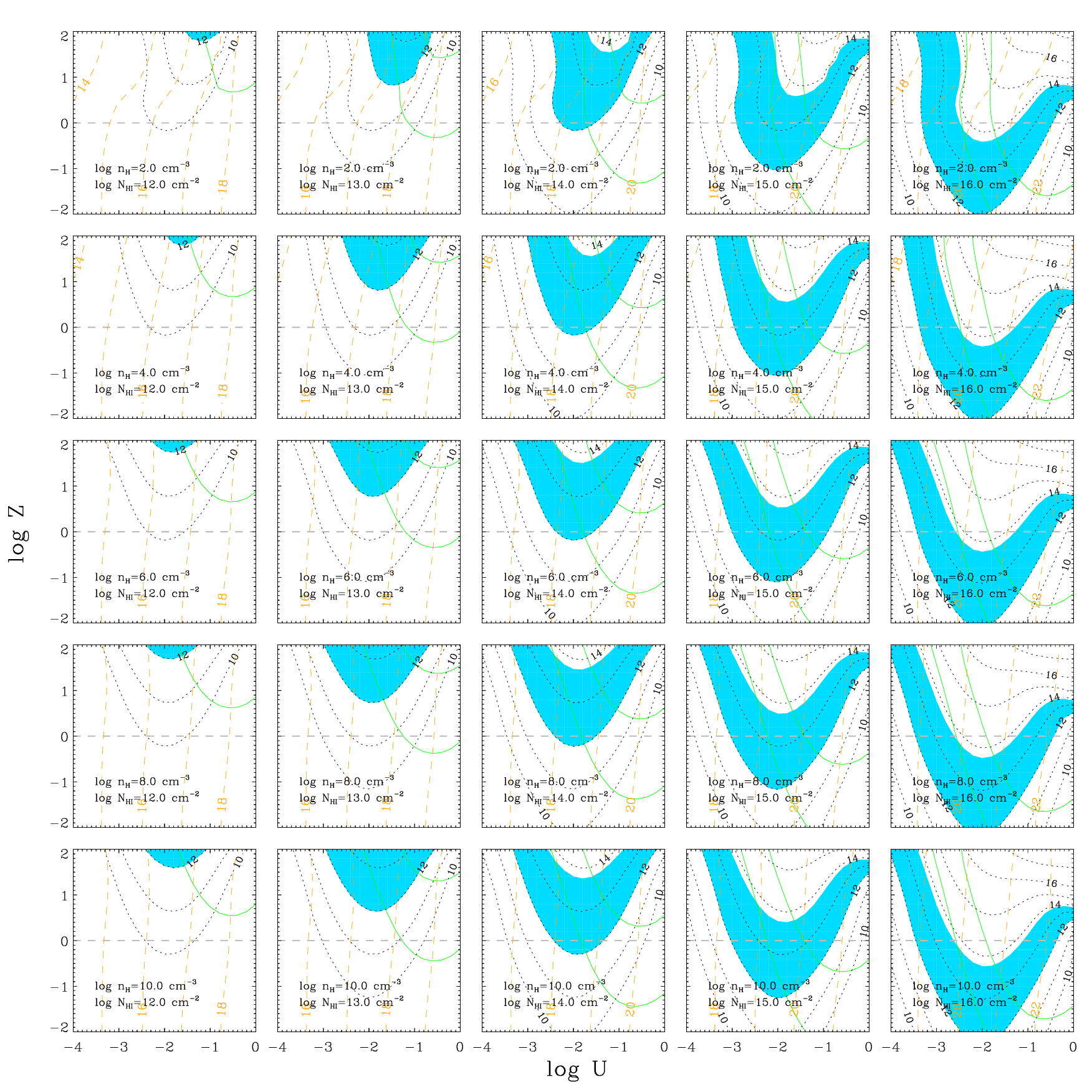}}
\caption{The simulated $N_{\rm Si\,IV}$, for different $N_{\rm H\,I}$
($10^{12}$, $10^{13}$, $10^{14}$, $10^{15}$, $10^{16}\,\rm cm^{-2}$) at each
{\nH} varied from ${\rm log}\,n_{\rm H}=2.0$ to $10\,\rm cm^{-3}$ by $2.0\,\rm dex$, 
are plotted as functions of $U$ and $Z$ by black dotted contours.
The cyan regions indicate the measurable regions of $N_{\rm Si\,IV}$ using 
{\SiIV} $\lambda1402.77$, assuming a Voigt velocity profile with a measured 
average FWHM of $17.2\,\rm km\,s^{-1}$ for our absorption components. These 
ranges are considered to be sensitive in measuring its column density as long as
the optical depth at the line center is in the range of $0.05$--$3$. Similarly, the two green
contours show the measurable ranges for $N_{\rm N\,V}$ using {\NV} $\lambda1238.82$,
with a measured average FWHM of $22.6\,\rm km\,s^{-1}$. The {\NH} for each
panel is exhibited in orange contours.
}
\label{figHISiIV}
\end{figure*}

As shown in Figure \ref{figmodT1}, the column density of {\SiIV} is sensitive to
metallicities at the condition when the modeled $N_{\rm H\,I}$ reaches the 
observed value. We would like to further explore the diagnostic power of {\SiIV} 
to the abundances by checking how $N_{\rm Si\,IV}$ varies with metallicity for
different {\nH} and $N_{\rm H\,I}$ values. We carried out comprehensive 
photoionization simulations to do the investigation.

\begin{figure}[htb]
\center{\includegraphics[width=8.4cm]  {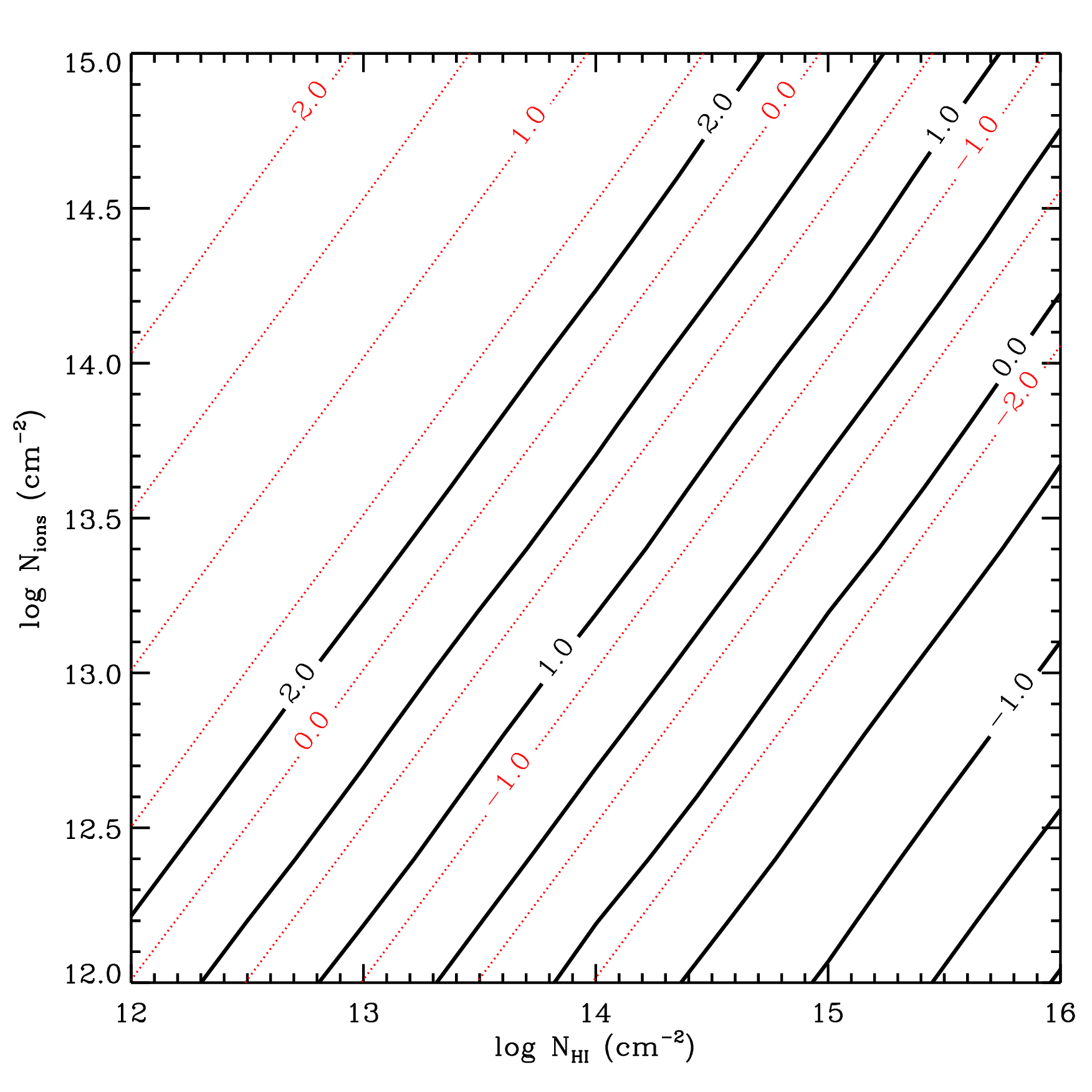}}
\caption{The minimum values of metallicity ($Z_{\rm min}$), suggested
by $N_{\rm H\,I}$ and $N_{\rm Si\,IV}$ at ${\rm log}\,n_{\rm H}=4\,{\rm cm^{-3}}$
(see Figure \ref{figHISiIV}), are plotted as functions of $N_{\rm H\,I}$ and
$N_{\rm Si\,IV}$ by the solid black line. Meanwhile, the $Z_{\rm min}$ suggested
by $N_{\rm H\,I}$ and $N_{\rm N\,V}$ is also exhibited, as shown by red dotted
lines for comparison.
}
\label{figminz}
\end{figure}

We assume a slab-shaped, homogeneous gas with 4B SED (see the details in Section
\ref{secnH}). The simulations were run over $-4\leqslant{\rm log }\,U\leqslant0$
and $-2\leqslant{\rm log}\,Z\leqslant2$, for a given {\nH}, at the condition
when the predicted $N_{\rm H\,I}$ reached a given value (e.g., observed value).
The $N_{\rm Si\,IV}$ contour distributions for different $N_{\rm H\,I}$
($10^{12}$, $10^{ 13}$, $10^{14}$, $10^{15}$, and $10^{16}\,{\rm cm^{-2}}$)
at each {\nH} varied from ${\rm log}\,n_{\rm H}=2.0$ to $10.0\,{\rm cm^{-3}}$
by $2.0\rm\,dex$ are plotted as black dashed lines in Figure \ref{figHISiIV}. Also
included in this figure are the {\NH} distribution contours (orange dashed lines).
We indicated the measurable $N_{\rm Si\,IV}$ and $N_{\rm N\,V}$
ranges,\footnote {When the optical depth at the deepest point
of the {\SiIV} absorption trough is within the range of 0.05--3 (in this case,
the corresponding residual flux ($e^{-\tau}$) is in the range of 0.95--0.05, and
the line is neither too weak nor severely saturated \citep{Zhou2019}), we think
that the {\SiIV} absorption lines can be detected and its column density can also
be measured. Assuming a Voigt velocity profile with $\rm FWHM=17.2\,{\rm km\,s^{-1}}$,
which is the average width of the {\SiIV} absorption line for the components
in SDSS J1220+0923,
for the {\SiIV} $\lambda1402.77$ absorption line, the measurable $N_{\rm Si\,IV}$ is therefore
$9.89\times10^{11}\,{\rm{cm}}^{-2}\leqslant N_{\rm Si\,IV}\leqslant5.78\times10^{13}\,{\rm cm}^{-2}$,
indicating that the $N_{\rm Si\,IV}$ can be measured over about 2 orders of
magnitude. Via similar analysis, the $N_{\rm N\,V}$, with the FWHM to the
measured average value ($22.6\,\rm km\,s^{-1}$) for the {\NV} $\lambda1238.82$, can be estimated to be in
the range of $4.81\times10^{12}\,{\rm cm^{-2}}\leqslant N_{\rm N\,V}\leqslant2.81\times10^{14}\,{\rm cm^{-2}}$.
}
obtained by constructing Voigt  profiles for the optical depths of the absorption
lines whose line widths adopt the observed values for
the absorption components in SDSS J1220+0923, as cyan regions. It can be seen clearly that the $N_{\rm Si\,IV}$ contour
distributions are almost the same for different {\nH} that varied more than
10 orders of magnitude. $N_{\rm Si\,IV}$ is very sensitive to $Z$ when the
$N_{\rm H\,I}$ is relatively low at ${\rm log}\,N_{\rm H\,I}<16.5\,{\rm cm^{-2}}$,
indicating that the metallicity of the absorbing gas can be diagnosed by
combining the information of $N_{\rm H\,I}$ and $N_{\rm Si\,IV}$.

The minimum values of metallicity ($Z_{\rm min}$), suggested by $N_{\rm H\,I}$
and $N_{\rm Si\,IV}$ at ${\rm log}\,n_{\rm H}=4.0\,{\rm cm^{-3}}$ (see Figure
\ref{figHISiIV}), were shown as functions of $N_{\rm H\,I}$ and $N_{\rm Si\,IV}$
in Figure \ref{figminz}. Also included in this figure are the $Z_{\rm min}$
values suggested by $N_{\rm H\,I}$ and $N_{\rm N\,V}$ at the same {\nH}. It can
be seen that for the same value of $N_{\rm Si\,IV}$ and $N_{\rm N\,V}$ for a
given $N_{\rm H\,I}$, $N_{\rm Si\,IV}$ leads to a $Z_{\rm min}$ that is
significantly higher than that derived from $N_{\rm N\,V}$ by approximately
1.8 dex. It is found that $Z_{\rm min}$, suggested by $N_{\rm H\,I}$ and
$N_{\rm Si\,IV}$, can be well described by the following equation for
$N_{\rm H\,I}\lesssim10^{16.5}\,\rm cm^{-2}$:
\begin{equation}\label{eqzmin}
{\rm log}\,Z_{\rm min}=-0.994\,{\rm log}\,N_{\rm H\,I}+0.940\,{\rm log}\,N_{\rm Si\,IV}+2.537,
\end{equation}
indicating that ${\rm log}\,Z_{\rm min}$ increases linearly with the increase
of ${\rm log}\,N_{\rm Si\,IV}$ for a given $N_{\rm H\,I}$, and decreases
linearly with the increase of ${\rm log}\,N_{\rm H\,I}$ for a given
$N_{\rm Si\,IV}$. It can be seen that the absorbing gas will have a relatively
large $Z_{\rm min}$ when the $N_{\rm H\,I}$ is relatively small while the
$N_{\rm Si\,IV}$ is relatively large. This phenomenon provides us with a way
to search for absorbing gas with high metallicities.

We use the measured total $N_{\rm H\,I}$ and total $N_{\rm Si\,IV}$ (see 
Table \ref{tabNcol}) to estimate the lower limit of the metallicity for 
the absorber as a whole in SDSS J1220+0923, and obtain its $Z_{\rm min}=2.49\,Z_{\sun}$, 
which is consistent with the estimated result of $\sim3.54Z_\sun$ in Section \ref{secUnN}. 
The $Z_{\rm min}$ for the seven components, determined using Equation (\ref{eqzmin}), 
are $1.50\,Z_{\sun}$ ($T1$), $2.08\,Z_{\sun}$ ($T2$), $3.08\,Z_{\sun}$ ($T3$), 
$2.12\,Z_{\sun}$ ($T4$), $2.25\,Z_{\sun}$ ($T5$), $5.76\,Z_{\sun}$ ($T6$), 
and $3.77\,Z_{\sun}$ ($T7$), all having supersolar metallicities. In particular, 
the component $T6$, with its $N_{\rm H\,I}$ exceeding $N_{\rm Si\,IV}$ by
only $1.101\,\rm dex$, exhibits metal abundances as high as $5.76\,Z_{\sun}$.
These estimated lower limits of metallicity are summarized in Table \ref{tabNcol}.

\subsection{Nonsolar [N/Si] and [C/Si]?}\label{secnonsolar}

\begin{figure}[htb]
\center{\includegraphics[width=8.4cm]  {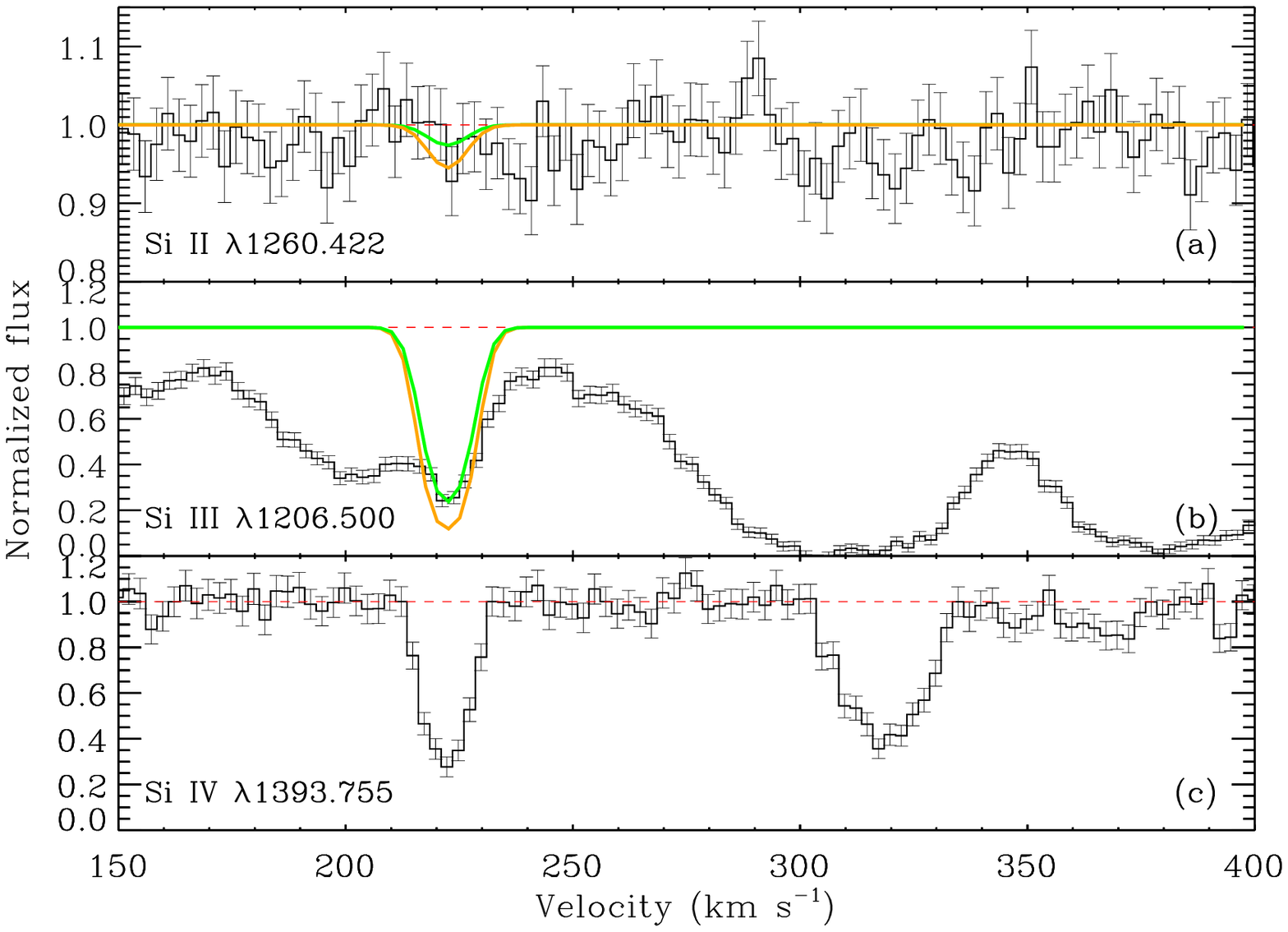}}
\caption{The normalized spectra for {\SiII} $\lambda1260.42$, {\SiIII} $\lambda1206.50$,
and {\SiIV} $\lambda1402.77$, respectively. The orange and green lines are the modeled
absorption troughs of {\SiII} $\lambda1260.422$ (panel (a)) and
{\SiIII} $\lambda1260.500$ (panel (b)), with the column densities predicted by
the best model using scaled solar abundance and reduced nitrogen abundance,
assuming their optical-depth profiles are identical to that of {\SiIV}.
}
\label{figSiIII}
\end{figure}

The column densities of {\NIII} predicted by the best model of component
$T1$, as shown in Section \ref{secUnN}, is ${\rm log}\,N_{\rm N\,III}=13.77\pm0.03\,{\rm cm^{-2}}$.
This is larger than that of the measured result (see Table \ref{tabNcol})
by $0.3\rm\,dex$. This difference hints that the nitrogen abundance may
deviate from that of the solar value. To illustrate this idea further, we plot a series
of figures, similar to Figure \ref{figmodT1}, adopting scaled solar abundance
except nitrogen. The results
show that with the decrease in the abundance of nitrogen, the locus of the measured
$N_{\rm Si\,IV}$ in the parameter space almost remains unchanged, and the locus
of measured $N_{\rm N\,V}$ only moves slightly in the direction of larger $U$,
while the predicted  $N_{\rm N\,III}$ gradually decreases. When the abundance
of nitrogen is 0.4 times that of the scaled solar value, the predicted $N_{\rm N\,III}$
matches that of the measured result. In this case, the best model results
are ${\rm log}\,U=-1.705\pm0.025$, ${\rm log}\,Z=0.200\pm0.025$,
${\rm log}\,N_{\rm H}=18.30\pm0.03\,\rm cm^{-2}$, and
${\rm log}\,n_{\rm H}=2.205\pm0.005\,{\rm cm^{-3}}$. The $U$, $Z$, and {\NH}
are only slightly larger than those estimated from the best-fit scaled solar abundance
modeling, as shown in Section \ref{secUnN}, while the
{\nH} is almost the same. The estimated distance is $12.3\pm0.5\,\rm kpc$,
which is similar to the result from the scaled solar abundance modeling
($14.8\pm0.5\,\rm kpc$, Section \ref{secUnN}).

Similarly, the best model for component $T1$ predicts $N_{\rm C\,III}$
and $N_{\rm C\,IV}$ to be larger than the observed values, as detailed
in Section \ref{secUnN}. This discrepancy also suggests that the carbon
abundance may deviate from the scaled solar abundance. Specifically,
when the carbon abundance is reduced to 0.13 times that of the
scaled solar value, the predicted $N_{\rm C\,III}$ and $N_{\rm C\,IV}$
are in agreement with the observations.

The ${\rm log}\,N_{\rm Si\,III}$ of component $T1$ predicted by the best model
using scaled solar abundance (Figure \ref{figmodT1}) and nonsolar N and C abundance
are $12.68\pm0.04$ and $12.50\pm0.04\rm\, cm^{-2}$, respectively. Meanwhile,
the corresponding ${\rm log}\,N_{\rm Si\,II}$ predicted are $11.18\pm0.08$ and
$10.85\pm0.08\,{\rm cm^{-2}}$. Assuming the optical depth of {\SiII} and
{\SiIII} have the same profiles as that of {\SiIV}, we then reproduced the
modeled absorption troughs of {\SiII} $\lambda1260.422$ and {\SiIII} $\lambda1206.500$,
using the column densities predicted by the best model with the scaled solar abundance
and nonsolar N and C abundances, as the orange and green lines shown in Figure
\ref{figSiIII}. It can be seen that the scaled solar abundance modeling is a
little deeper than the observed trough, while the nonsolar N and C abundance
modeling agrees with the observations. These experiments are all consistent
with our speculation that the nitrogen and carbon abundance
in the absorber might be lower than that of the solar values.
Similar phenomena have also been reported in the literature
\citep[e.g.,][]{Zafar2014}.

\subsection{{\Lya} and {\SiIV} sample}\label{secsample}
As we have demonstrated in Figures \ref{figHISiIV} and \ref{figminz}, the
combined information of $N_{\rm H\,I}$ and $N_{\rm Si\,IV}$ can give
indicative clues on the metallicities of the absorbers in quasars.
We can use this idea to systematically search for quasars that may
show high abundances. We search for the absorption-line quasars
that show detectable {\Lya}, {\SiIV} and {\NV} absorption troughs.
We focused on the intrinsic absorbers, and the selection criterion is
based on the detectability of high-ionization {\NV} absorption lines,
as reported by S. Perrotta et al. (\citeyear{Perrotta2016}). To avoid
contamination from the {\Lya} forests on {\NV} absorption lines,
we only selected absorption-line systems with velocities of $v>-5000\,\rm km\,s^{-1}$.
In addition, three quasars that had been checked before, having
velocities outside of the above criteria,  were also included in our sample.
The details of the sample selection for 28 absorption-line systems in 25 quasars
are presented in Appendix \ref{appendix_B}.

\begin{figure}[htb]
\center{\includegraphics[width=8.4cm]  {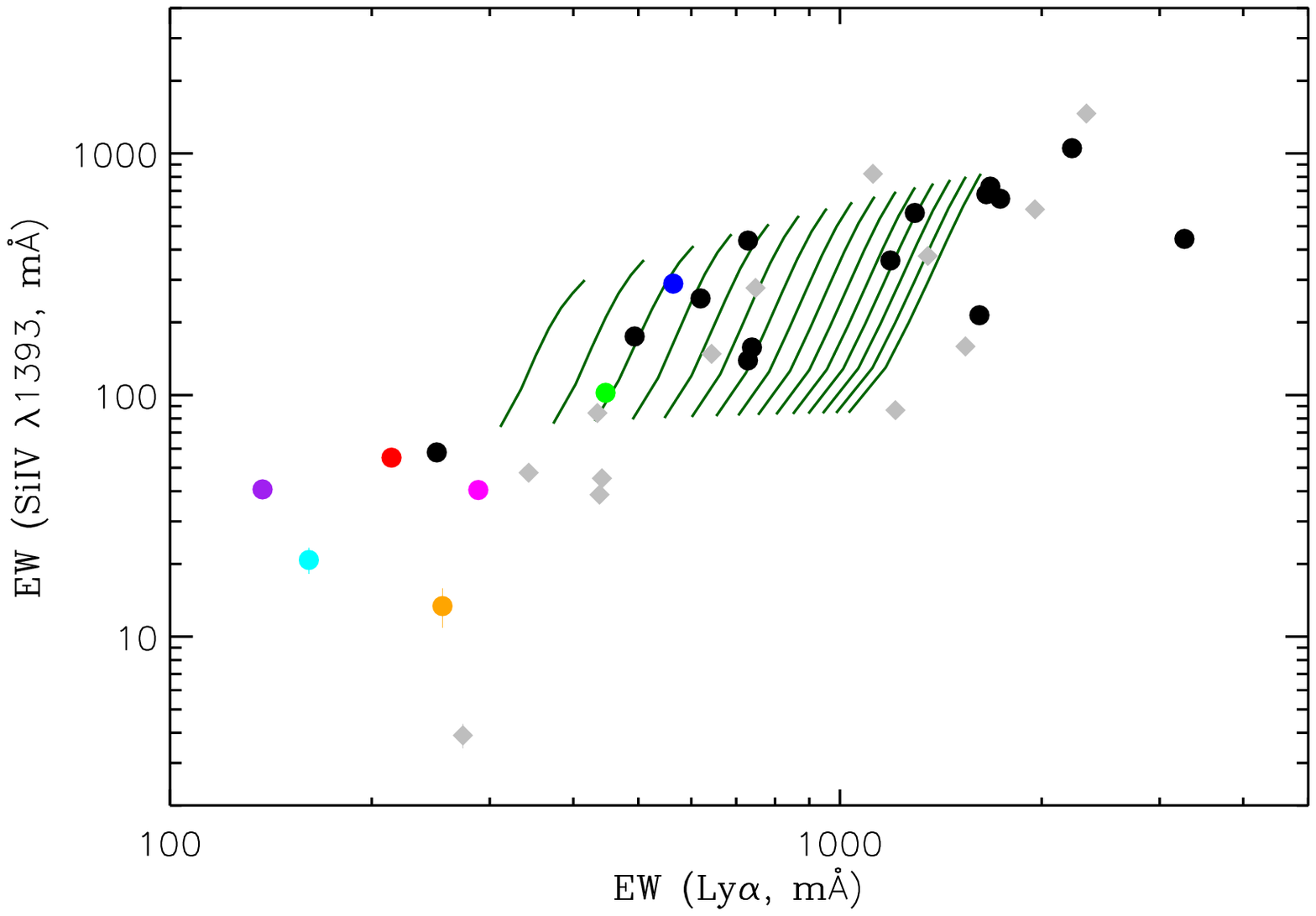}}
\caption{The measured EWs of {\SiIV} $\lambda1393.76$ presented as a function
of those of {\Lya} for a quasar spectrum sample selected from VLT/UVES and
Keck/HIRES spectra, based on the criteria that (1) velocities $> \rm -5000\,km\,s^{-1}$,
and (2) {\NV} absorptions are detected. The gray-filled diamonds represent the inflows
with red-shifted velocities $>164\,\rm km\,s^{-1}$, while the black solid
circles represent absorption lines with blue-shifted velocities.
The EWs for {\HI} larger than $5000\,\rm m${\AA} are not presented for the
sake of clarity. The red-, green-, blue-, orange-, magenta-, purple-, and cyan-filled
circles are for components $T1$ to $T7$ of SDSS J1220+0923. The dark
green lines indicate $Z_{\rm min}=Z_\sun$, plotted from Equation (\ref{eqzmin}),
where $N_{\rm H\,I}$ and $N_{\rm Si\,IV}$ are converted from the corresponding
EWs with $b=20$--$90\,\rm km\,s^{-1}$ in a step by $5\,\rm km\,s^{-1}$ from
left to right.
}
\label{figsampl}
\end{figure}

We plotted the measured equivalent widths (EWs) of {\SiIV} $\lambda1393.76$
versus those of {\Lya}, from Table \ref{tabsampl} in Appdix \ref{appendix_B},
in Figure \ref{figsampl}. The red-, green-, blue-, orange-,
magenta-, purple-, and cyan-filled circles represent the absorption components
from $T1$ to $T7$ in SDSS J1220+0923. As shown in Figure \ref{figminz},
the absorber's $Z_{\rm min}$ can be estimated by $N_{\rm H\,I}$ and $N_{\rm Si\,IV}$.
Using the COG for a given $b$ value, the ionic column densities can
be derived from the EWs of the absorbing lines, as illustrated in Figure \ref{figHI}. We
overplotted the $Z_{\rm min}=Z_{\sun}$ contour in Figure \ref{figminz} to Figure
\ref{figsampl} for $b$ values ranging from 20 to $90\,{\rm km\,s^{-1}}$ in a
step of $5\,{\rm km\,s^{-1}}$, as shown by dark green lines. It can be seen that
the absorbing components in SDSS J1220+0923 have small $\rm EW_{\rm Ly\alpha}$, reinforcing our result that they
have supersolar metallicities.

There are 12 absorption-line systems in 12 quasars that exhibit red shifted absorption
lines, with velocities $>164$ $\rm km\,s^{-1}$, as presented in Table \ref{tabsampl}.
The measured EWs of {\Lya} and {\SiIV}, as shown in Figure \ref{figsampl}
(gray-filled diamonds), imply the inflowing gas in these quasars might have high
metal abundances. These objects will be further analyzed in the future.

\section{Summary, Implications, and future Works} \label{secsumm}
In summary, the high-redshift quasar SDSS J1220+0923 exhibits abundant absorption
lines, which are close to the emission redshift of the quasar, with red-shifted
velocity separation from the quasar systemic redshift ranging from 200 to $900\,\rm km\,s^{-1}$. 
These include high-ionization species of {\NV}, {\CIV}, and {\SiIV},
low-ionization species of {\NIII}, \ion{C}{3}, and {\SiIII}, and
absorption lines from {\HI} Lyman series. The properties of the absorber
were probed by jointly using absorption lines and photoionization simulations.
The relationships between {\nH} and ($U$ and $Z$), and between {\NH} and
($U$ and $Z$) were established based on the measured column densities of
{\NIII}, {\NIII*}, and {\HI}. Thus, the four parameters ($U$, {\nH}, {\NH} and $Z$)
of the absorbing gas are determined in a 2D parameter space. We found that
the component $T1$ of the absorber has a metallicity of $\sim1.54\,Z_\sun$,
and the absorber as a whole was determined to be $\sim3.54\,Z_\sun$. Meanwhile,
the lower limits of metallicities of the other six components were also determined, 
and all exhibit supersolar values. It is intriguing to detect such a high-abundance 
inflow of gas in a quasar with a redshift of $z\sim 3$ when the Universe is still young.
The metal-strong inflowing gas, located at $\sim15\rm kpc$, is most likely originated 
in situ and driven by stellar processes.

An empirical formula for estimating the abundance lower limit of absorbers, using
the measured $N_{\rm H\,I}$ and $N_{\rm Si\,IV}$, has been proposed. It suggests
that metal-strong absorbers, such as remnants of supernovae, could be selected by
relatively weak {\HI} Lyman series and strong {\SiIV} absorption lines.

A systematic study of the metallicities for quasar absorbers can be conducted in 
the future with methods as follows: (1) search for quasar spectra where absorption 
lines from {\HI} Lyman series and {\NV} doublet can be detected, indicating that 
they are likely intrinsic absorbers \citep{Perrotta2016}; (2) establish the relationship 
between {\nH} and ($U$ and $Z$) using the {\HI} absorption lines and low-ionization 
absorption lines from the ground and excited states, such as {\SiII}, {\CII}, and {\NIII}; 
(3) high-ionization absorption lines such as {\SiIV}, {\CIV}, and {\NV} can be used 
to constrain metal abundance in the ($U$ and $Z$) parameter space.

The authors thank the anonymous reviewer for the constructive suggestions.
This work was jointly supported by the National Key R\&D Program of China
(grant Nos. 2022YFC2807300 and 2022YFF0503402), grant No. KJSP2020010102, the National Natural Science Foundation of
China (NSFC; Nos. 11503023, and 12233005), and the Shanghai Natural Science Foundation (grant Nos.
19ZR1462500, 14ZR1444100, 20ZR1473600, 22ZR1481200, 21ZR1469800, and 21ZR1474200).
This work has made use of data products from SDSS,
UKIDSS, Catalina Sky Survey, data obtained through the UVES Paranal Observatory
Project (ESO DDT Program ID 266.D-5655), and through the Telescope Access Program
(TAP), which has been funded by the Strategic Priority Research Program The
Emergence of Cosmological Structures (grant No. XDB09000000), National
Astronomical Observatories, Chinese Academy of Sciences, and the Special Fund for
Astronomy from the Ministry of Finance.

\appendix
\section{A. Simulation of $N_{\rm H\,I}$ Ranges Using Lyman Absorption lines and the Lyman-limit System} \label{appendix_A}
\begin{figure*}[htb]
\center{\includegraphics[width=18cm]  {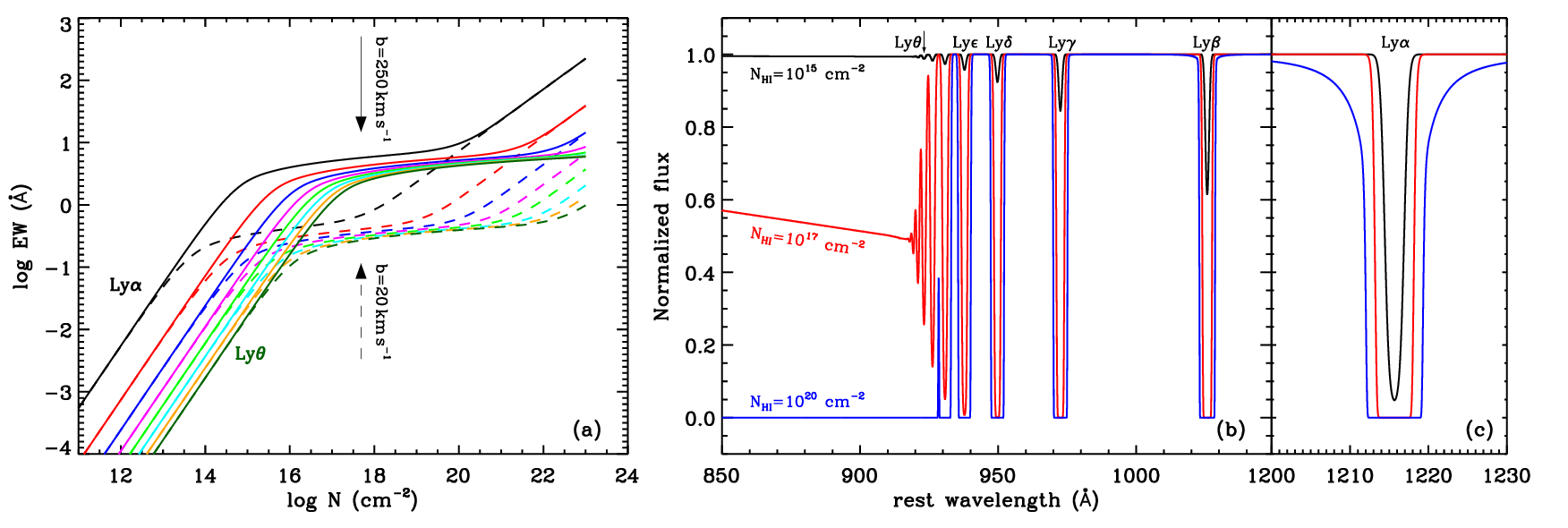}}
\caption{(a): COGs of absorption lines from Lyman series ({\Lya}, {\Lyb} up to Ly$\theta$)
with $b=250\,\rm km\,s^{-1}$ (solid lines) and $b=20\,\rm km\,s^{-1}$ (dashed lines).
(b), (c): the simulated Lyman absorption lines and the Lyman-limit system, for
$b=250\,\rm km\,s^{-1}$ at $N_{\rm H\,I}=10^{15}$, $10^{17}$, and $10^{20}\,\rm cm^{-2}$,
which are shown as black, red, and blue lines, respectively.
}
\label{figHI}
\end{figure*}

Figure \ref{figHI} (a) shows the COGs of {\HI} Lyman series,
{\Lya}, {\Lyb}, up to Ly$\theta$, with $b=250$ and $20\,\rm km\,s^{-1}$. These
two $b$ values correspond to the absorption width of
mini-BALs ($b=250\,\rm km\,s^{-1}$) and the typical {\Lya} forest ($b=20\,\rm km\,s^{-1}$)
\citep{Hu1995,Lu1996,Kim1997}, respectively. The COGs
are divided into three parts: `linear' part, `flat' part, and `damping' part
\citep{Peterson1997}. We plot the simulated Lyman absorption lines and the Lyman-limit
system for $b=250\,\rm km\,s^{-1}$ at typical $N_{\rm H\,I}$ values,
corresponding to the three parts, in Figures \ref{figHI} (b) and (c). To evaluate the
$N_{\rm H\,I}$ ranges measured by Lyman absorption lines and the Lyman-limit system,
we considered that they can be used to measure $N_{\rm H\,I}$ values when their
residual flux is within the range of 0.95--0.05, in which they are neither too weak
nor severely saturated \citep{Zhou2019}. For $b=250\,\rm km\,s^{-1}$, in the linear
part, the lower limit of $N_{\rm H\,I}$, estimated using the {\Lya} absorption line
when the residual flux at the deepest point is 0.95 (0.05 for optical depth), is
about $10^{13.4}\,\rm cm^{-2}$. Meanwhile, the upper limits of $N_{\rm H\,I}$
measured using {\Lya} and Ly$\theta$ when the residual flux at the deepest point
is 0.05, are about $10^{15.8}$ and $10^{17.5}\,\rm cm^{-2}$.
The Lyman absorption lines with wavelengths less than those of Ly$\theta$ are heavily
blended (see Figure \ref{figHI} (b)). Thus, for $b=250\,\rm km\,s^{-1}$in the
linear part, the measured $N_{\rm H\,I}$ range is about $10^{13.4}$--$10^{17.5}\,\rm cm^{-2}$.
At the same time, the estimated $N_{\rm H\,I}$ range that can be measured, in
the linear part for $b=20\,\rm km\,s^{-1}$, is about $10^{12.3}$ (by {\Lya}) to
$10^{16.4}\rm\,cm^{-2}$ (by Ly$\theta$).

The optical depth of the Lyman-limit system for $\lambda\leqslant912\,${\AA} is \citep{Mo2010},
\setcounter{equation}{0}
\renewcommand\theequation{A\arabic{equation}}
\begin{equation}
\tau\left(\lambda\right)=\left(\frac{N_{\rm H\,I}}{1.62\times10^{17}\,\rm cm^{-2}}\right)\left(\frac{\lambda}{912\,{\mathring{\rm A}}}\right)^3.
\end{equation}
When the residual flux at $912\,${\AA} is 0.95, the $N_{\rm H\,I}$ is
$10^{15.9}\,\rm cm^{-2}$, which is considered the lower limit of
$N_{\rm H\,I}$ that can be measured by the Lyman-limit system. The Lyman-limit
system is powerful for measuring $N_{\rm H\,I}$, as indicated by the
measurable optical depth at specific wavelengths. For instance, when $\tau$
is 3 at $\lambda$ of $714\,{\mathring{\rm A}}$, with a residual flux of 0.95,
the corresponding $N_{\rm H\,I}$ measured by the Lyman-limit system is
$10^{18}\,\rm cm^{-2}$. For higher values of $N_{\rm H\,I}$, it
can also be measured by using a damped wing of {\Lya} absorptions (see Figure
\ref{figHI} (c) for $N_{\rm H\,I}=10^{20}\,\rm cm^{-2}$ as an example). Thus,
it can be suggested that measuring $N_{\rm H\,I}$ is feasible in a remarkably wide range,
from approximately $10^{13}$ to more than $10^{20}\rm\, cm^{-2}$, provided
that the spectra are of a sufficiently high resolution and spectral S/N.

\section{B. {\Lya} and {\SiIV} sample selection} \label{appendix_B}
\begin{deluxetable*}{llrrrrrrccc}
\tabletypesize{\scriptsize}
\tablewidth{0pt}
\tablecaption{{\Lya} and {\SiIV} sample}
\tablehead{
		\colhead {Source}           &
		\colhead {$z_{\rm em}$}           &
		\colhead {$v_{\rm ave}$\tablenotemark{a}}          &
		\colhead {EW ({\Lya}) }           &
		\colhead {EW (\ion{Si}{4})}           &
		\colhead {EW ({\ion{Si}{4}})}           &
		\colhead {EW ({\ion{N}{5}})}           &
		\colhead {EW ({\ion{N}{5}})}           &
		\colhead {Sample}           & \\
		\colhead {}           &
		\colhead {}           &
		\colhead {($\rm km~s^{-1}$)}           &
		\colhead {(m$\mathring{\rm A}$)}           &
		\colhead {($\lambda1393$, m$\mathring{\rm A}$)}           &
		\colhead {($\lambda1402$, m$\mathring{\rm A}$)}           &
		\colhead {($\lambda1238$, m$\mathring{\rm A}$)}           &
		\colhead {($\lambda1242$, m$\mathring{\rm A}$)}           &
				}
\startdata
J001602.40-001225.1 & $2.0918\pm0.0002$\tablenotemark{b} & $-6161.40$& $1189.3\pm1.4$     & $360.6\pm3.9$ & $219.3\pm4.3$ & $<219.6\pm0.6$ & $75.6\pm0.8$ &  SQUAD\\
J004131.43-493611.5 & $3.2331\pm0.0009$\tablenotemark{c} & $-85.10$  & $1615.1\pm1.8$     & $214.0\pm1.4$ & $185.5\pm1.2$ & $116.2\pm2.0$ & $64.2\pm2.2$  & SQUAD\\
J010604.41-254651.3 & $3.3794\pm0.0002$\tablenotemark{b} & $-1271.40$& $3268.1\pm2.1$     & $442.9\pm6.9$ & $315.6\pm6.1$ & $621.1\pm4.8$ & $428.1\pm5.1$  & SQUAD\\
J010821.72+062327.1 & $1.9692\pm0.0001$\tablenotemark{b} & $-3484.46$& $1734.8\pm2.2$     & $649.7\pm3.3$ & $549.8\pm3.2$ & $167.8\pm3.4$ & $101.8\pm3.4$   &SQUAD\\

J012227.89-042127.1 & $1.9703\pm0.0002$\tablenotemark{d} & $-613.53$ & $1676.7\pm21.2$    & $729.1\pm7.4$ & $635.2\pm8.8$ & $1454.0\pm17.6$ & $1564.6\pm20.4$  & SQUAD\\
J012227.89-042127.1 & $1.9703\pm0.0002$\tablenotemark{d} & $164.53$  & $1539.5\pm18.3$    & $159.0\pm9.4$ & $131.5\pm8.8$ & $<911.3\pm15.0$ & $438.2\pm20.2$  & SQUAD\\
J013857.44-225447.3 & $1.8961\pm0.0001$\tablenotemark{c} & $255.49$  & $1954.8\pm1.5$     & $587.1\pm5.0$ & $474.5\pm5.1$ & $679.5\pm4.9$ & $504.3\pm5.6$   &SQUAD\\
J022620.49-285750.7 & $2.1750\pm0.0001$\tablenotemark{e} & $-189.29$ & $250.1\pm0.8$      & $57.9\pm1.8$ & $42.5\pm1.8$ & $165.7\pm1.5$ & $127.4\pm1.5$  & SQUAD\\
J030640.90-301031.6 & $2.0997\pm0.0005$\tablenotemark{c} & $-8346.26$& $1293.3\pm16.5$    & $567.3\pm19.2$ & $432.6\pm21.4$ & $415.2\pm8.3$ & $336.2\pm8.5$  & SQUAD\\
J030643.75-301107.7 & $2.1191\pm0.0001$\tablenotemark{c} & $-597.44$ & $2220.0\pm5.6$     & $1051.5\pm48.3$ & $683.8\pm41.8$ & $1630.1\pm12.4$ & $1591.4\pm14.1$  & SQUAD\\
J030643.75-301107.7 & $2.1191\pm0.0001$\tablenotemark{c} & $-70.64$  & $738.9\pm3.8$      & $157.4\pm23.0$ & $272.6\pm21.2$ & $214.6\pm6.9$ & $202.5\pm17.0$  & SQUAD\\
J031006.09-192124.0 & $2.1434\pm0.0001$\tablenotemark{e} & $-1947.55$& $728.8\pm2.2$      & $139.1\pm3.3$ & $145.9\pm2.9$ & $163.5\pm1.6$ & $141.2\pm2.3$  & SQUAD\\
J031009.09-192207.2 & $2.1239\pm0.0001$\tablenotemark{e} & $-176.97$ & $619.3\pm1.7$      & $251.2\pm4.3$ & $220.8\pm4.2$ & $74.2\pm2.0$ & $35.1\pm3.4$   &SQUAD\\
J040356.64-170321.8 & $4.2423\pm0.0001$\tablenotemark{b} & $-670.19$ & $729.1\pm1.1$      & $<435.9\pm4.0$ & $<361.7\pm4.7$ & $51.6\pm1.8$ & $32.9\pm1.8$  & SQUAD\\
J042707.29-130253.6 & $2.1652\pm0.0001$\tablenotemark{c} & $731.73$  & $343.0\pm1.5$      & $47.7\pm3.1$ & $32.8\pm3.2$ & $39.2\pm1.9$ & $29.2\pm2.2$  & SQUAD\\
J044017.16-433308.5 & $2.8582\pm0.0001$\tablenotemark{c} & $673.15$  & $1210.1\pm2.5$     & $86.6\pm5.2$ & $48.3\pm5.3$ & $397.9\pm3.2$ & $382.8\pm3.6$  & SQUAD\\
J044821.74+095051.4 & $2.1093\pm0.0002$\tablenotemark{c} & $-350.20$ & $1654.1\pm10.5$    & $678.2\pm18.4$ & $656.7\pm17.9$ & $1507.4\pm12.9$ & $1449.0\pm10.3$  & SQUAD\\
J044821.74+095051.4 & $2.1093\pm0.0002$\tablenotemark{c} & $348.42$  & $2333.1\pm13.9$    & $1463.7\pm24.9$ & $1661.7\pm19.7$ & $<1779.6\pm14.0$ & $1000.6\pm19.7$ &SQUAD\\
J053007.95-250329.7 & 2.813\tablenotemark{h}             & $-153.99$ & $22508.4\pm6.2$    & $821.8\pm0.8$ & $625.4\pm0.9$ & $184.4\pm1.5$ & $94.3\pm1.5$  & SQUAD  \\
J100731.40-333305.7 & $1.8362\pm0.0005$\tablenotemark{c} & $222.17$  & $1120.2\pm33.6$    & $824.0\pm21.0$ & $760.6\pm21.9$ & $917.4\pm37.2$ & $883.9\pm43.8$ & SQUAD\\
J234628.26+124858.1\tablenotemark{f} & 2.515             & $4624.79$ & $441.2\pm13.2$     & $45.2\pm1.7$ & $38.2\pm1.8$ & $191.3\pm2.2$ & $121.9\pm2.6$ & SQUAD \\
J234819.19+005717.5 & $2.1551\pm0.0002$\tablenotemark{c} & $1196.90$ & $434.4\pm3.4$      & $84.2\pm7.2$ & $48.0\pm7.4$ & $99.0\pm3.0$ & $87.9\pm3.9$ &SQUAD\\
J235034.25-432559.6 & $2.8893\pm0.0001$\tablenotemark{b} & $611.50$  & $273.6\pm0.2$      & $3.9\pm0.9$ & $2.5\pm0.9$ & $260.2\pm0.2$ & $201.5\pm0.3$ & SQUAD\\
J092759.26+154321.3 & $1.8073\pm0.0010$\tablenotemark{g} & $-8052.61$& $<14571.6\pm136.3$ & $423.9\pm13.0$ & $291.6\pm12.4$ & $391.1\pm13.6$ & $331.3\pm10.3$& KODIAQ\\
J093857.01+412821.2 & $1.9624\pm0.0003$\tablenotemark{g} & $-2485.81$& $493.7\pm5.6$      & $174.8\pm3.7$ & $114.7\pm4.2$ & $986.3\pm4.3$ & $913.2\pm5.9$ & KODIAQ   \\
J120147.91+120630.2 & $3.5219\pm0.0006$\tablenotemark{b} & $292.59$  & $1352.3\pm1.7$     & $376.5\pm4.4$ & $278.1\pm3.4$ & $223.4\pm3.0$ & $134.7\pm3.4$ & KODIAQ \\
J145435.18+094100.0 & $1.9515\pm0.0008$\tablenotemark{g} & $174.89$  & $<748.1\pm3.4$     & $277.6\pm6.5$ & $188.9\pm10.6$ & $<651.6\pm5.1$ & $<542.5\pm6.1$ & KODIAQ \\
J160455.40+381201.0 & $2.5371\pm0.0007$\tablenotemark{g} & $1456.64$ & $643.1\pm0.4$      & $148.1\pm0.0$ & $103.2\pm0.0$ & $16.3\pm0.0$ & $8.5\pm0.4$   & KODIAQ  \\
J234628.26+124858.1\tablenotemark{f}       & 2.515       & $4630.81$ & $437.7\pm0.0$      & $38.7\pm0.0$ & $23.6\pm0.0$ & $183.6\pm0.0$ & $116.8\pm0.0$  & KODIAQ
\enddata
\label{tabsampl}
\tablenotetext{a}{The mean velocity of the {\SiIV} $\lambda1393$ absorption trough weighted by its absorption depth.}
\tablenotetext{b}{The redshifts are determined using UVES spectra by the \ion{O}{1} $\lambda1302.168$ emission line.}
\tablenotetext{c}{The redshifts are determined using UVES spectra by the \ion{C}{3}] emission line.}
\tablenotetext{d}{The redshifts are determined using TripleSpec spectrum by the [\ion{O}{3}] emission line.}
\tablenotetext{e}{The redshifts are determined using UVES spectra by the \ion{Mg}{2} emission line.}
\tablenotetext{f}{This source has been observed by both UVES and KODIAQ, and the redshift, which is from the
SQUAD catalog \citep{Murphy2019}, might have considerable uncertainty since the emission lines are broad and
contaminated by Fe lines.}
\tablenotetext{g}{The redshifts are determined using SDSS spectra by the \ion{Mg}{2} emission line.}
\tablenotetext{h}{The redshifts are from the SQUAD catalog \citep{Murphy2019}, which might have considerable uncertainty since the emission lines are broad.}
\end{deluxetable*}

The sample is derived from two high-resolution quasar spectral data sets. The
first is the continuum-normalized spectra in the public data release of the
Keck Observatory Database of Ionized Absorption toward Quasars (KODIAQ)
including DR1 170 quasars) \citep{OMeara2015} and DR2 (130 quasars) \citep{OMeara2017},
at a redshift range of $0.07<z_{\rm e}<5.29$, observed with the High-Resolution Echelle
Spectrograph \citep[HIRES;][]{Vogt1994} at the Keck I telescope.
The second consists of 476 quasars with continuum-normalized spectra from the first data release of
SQUAD \citep{Murphy2019}, with a redshift range of $z_{\rm e}=0.01-6.31$, observed
with VLT/UVES \citep{Dekker2000}.

Our search focused on the so-called associated absorption lines \citep{Foltz1986,Hamann2004},
using the detectability of the {\NV} absorption doublet as a criterion, which
offers the best statistical tool to identify intrinsic systems, as reported by
S. Perrotta et al. (\citeyear{Perrotta2016}). We also rejected the objects where the
{\SiIV} doublet absorption troughs are heavily blended.

We calculated the rest-frame EWs for the absorption trough
of {\Lya}, {\SiIV} doublet, and {\NV} doublet for each absorption system in the
sample, and listed the results for 28 absorption systems in 25
quasars in Table \ref{tabsampl}. We have determined the systemic redshifts
for each source in the sample, employing emission lines in decreasing order
of accuracy: \ion{O}{1}, {\OIII}, {\MgII}, and {\CIII} \citep{Shen2016},
if they could be observed. The measured redshifts are detailed in Table \ref{tabsampl},
with the specific emission line used indicated. For sources in the first SQUAD data release,
we use the VLT/UVES spectra \citep{Dekker2000} to measure redshifts. On the other hand,
for the sources in the KODIAQ data release, we use the SDSS spectra, as only normalized
spectra have been released \citep{OMeara2015,OMeara2017}. It is worth noting that, for
one of the sources (J012227.89-042127.1), we also conducted NIR spectroscopic
observations on 2020 October 6, using the Hale 200 inch telescope with the TripleSpec
spectrograph, and the systemic redshift was measured using the {\OIII} $\lambda5008.24$
emission line. With the measured redshifts, the weight-averaged absorber
velocities are measured using the absorption profile of
{\SiIV} $\lambda1393.76$, as listed in Table \ref{tabsampl}.

\end{document}